\documentclass[fleqn,usenatbib]{mnras}

\usepackage{newtxtext,newtxmath}
\usepackage[T1]{fontenc}

\usepackage{graphicx}	% Including figure files
\usepackage{amsmath}	% Advanced maths commands
\usepackage[dvipsnames]{xcolor}
\usepackage{booktabs,array}
\usepackage{bm}

\title[Generalized Magnetar Model]{A Generalized Semi-Analytic Model for Magnetar-Driven Supernovae}

\author[Omand and Sarin]{
Conor M. B. Omand$^{1}$\thanks{E-mail: conor.omand@astro.su.se}
and Nikhil Sarin$^{2,3}$
\\
$^{1}$The Oskar Klein Centre, Department of Astronomy, Stockholm University, AlbaNova, SE-106 91 Stockholm, Sweden\\
$^{2}$Nordita,
Stockholm University and KTH Royal Institute of Technology
Hannes Alfvéns väg 12, SE-106 91 Stockholm, Sweden\\
$^{3}$The Oskar Klein Centre, Department of Physics, Stockholm University, AlbaNova, SE-106 91 Stockholm, Sweden\\
}

\pubyear{2023}

\begin{document}
\label{firstpage}
\pagerange{\pageref{firstpage}--\pageref{lastpage}}
\maketitle

\begin{abstract}
Several types of energetic supernovae, such as superluminous supernovae (SLSNe) and broad-line Ic supernovae (Ic-BL SNe), could be powered by the spin-down of a rapidly rotating magnetar. Currently, most models used to infer the parameters for potential magnetar-driven supernovae make several unsuitable assumptions that likely bias the estimated parameters. In this work, we present a new model for magnetar-driven supernovae that relaxes several of these assumptions and an inference workflow that enables accurate estimation of parameters from lightcurves of magnetar-driven supernovae.
In particular, in this model, we include the dynamical evolution of the ejecta, coupling it to the energy injected by the magnetar itself while also allowing for non-dipole spin down.  
% The main changes from previous models are the coupling of the magnetar energy injection to the kinetic energy of the supernova and the addition of the magnetar braking index as a free parameter, allowing the exploration of non-vacuum-dipole spin down. 
We show that the model can reproduce SLSN and Ic-BL SN light curves consistent with the parameter space from computationally expensive numerical simulations. We also show the results of parameter inference on four well-known example supernovae, demonstrating the model's effectiveness at capturing the considerable diversity in magnetar-driven supernova lightcurves.  
The model fits each light curve well and recovers parameters broadly consistent with previous works. 
This model will allow us to explore the full diversity of magnetar-driven supernovae under one theoretical framework, more accurately characterize these supernovae from only photometric data, and make more accurate predictions of future multiwavelength emission to test the magnetar-driven scenario better.
\end{abstract}


%%%%%%%%%%%%%%%%%%%%%%%%%%%%%%%%%%%%%%%%%%%%%%%%%%

%%%%%%%%%%%%%%%%% BODY OF PAPER %%%%%%%%%%%%%%%%%%

\section{Introduction}

Recent wide-field high-cadence surveys, such as the Zwicky Transient Facility \citep[ZTF,][]{ztf_paper} and the Asteroid Terrestrial-impact Last Alert System \citep[ATLAS,][]{Tonry2011, Tonry2018}, can discover more than 1000 supernovae per year \citep{Cappellaro2022}, and next generation facilities such as Rubin Observatory \citep{Ivezic2019} will be able to discover more than 1000 supernovae per night \citep{LSSTScienceCollaboration2009, Stritzinger2018}. Core-collapse supernovae (CCSNe), caused by the deaths of massive stars, tend to have explosion energies of $\sim$ 10$^{51}$ erg and radiated energies $\sim$ 10$^{49}$ erg, which can be well explained by models \citep{Arnett1980, Arnett1982}. However, several classes of CCSNe have energies higher than these standard models. Superluminous supernovae (SLSNe) radiate $\sim$ 10 $-$ 100 times more energy than a standard CCSN \citep{Gal-Yam2012, Nicholl2021} and broad-line Type Ic supernovae (SNe Ic-BL) have inferred kinetic energies $\sim$ 10 times higher than a typical CCSN \citep[e.g.][]{Taddia2019, Modjaz2019}. These energetic supernovae have both been associated with long- or ultra-long gamma-ray bursts (GRBs) \citep{Gendre2013, Nakauchi2013, Levan2014, Cano2017}, and SLSNe and SNe Ic-BL also have similar host galaxies \citep{Lunnan2014, Chen2015, Leloudas2015b, Angus2016, Chen2017, Schulze2018, Orum2020}, and have similar spectral features at both early \citep{Pastorello2010, Inserra2013, Nicholl2013, Liu2017, Blanchard2019} and late \citep{Milisavljevic2013,Jerkstrand2017,Nicholl2017} times. Other types of unusually energetic supernovae, such as fast blue optical transients (FBOTs) and bright ultra-stripped supernovae (USSNe), have been suggested to have similar power sources as SLSNe and SNe Ic-BL \citep{Liu2022, Sawada2022}.

Several different models can be used to explain the high energies of one or both of SLSNe or SNe Ic-BL.  Stars with masses M$_*$ $\gtrsim$ 130 M$_\odot$ can explode as pair instability supernovae (PISNe) \citep{Barkat1967, Heger2002, Gal-Yam2009}, which can generate tens of solar masses of $^{56}$Ni and present as an extremely long-lived, luminous supernovae.  Slightly less massive stars, with 130 M$_\odot$ $\gtrsim$ M$_*$ $\gtrsim$ 100 M$_\odot$, can eject shells of material through the pair-instability without being completely destabilized, and collisions between the ejected shells can be as luminous as an SLSN \citep{Heger2003, Woosley2007, Chatzopoulos2012, Yoshida2016, Woosley2017} - these are known as pulsational pair-instability supernovae (PPISNe). The collision of supernova ejecta and circumstellar material surrounding the progenitor, ejected through either a steady wind, eruptive mass loss, or binary interaction \citep{Smith2014}, can convert much of the supernova kinetic energy into radiation, leading to a highly luminous supernova \citep{Chatzopoulos2012, Chatzopoulos2013, Villar2017, Jiang2020}.  Finally, the compact remnant can inject some energy into the ejecta; this energy may come from fallback accretion onto a central black hole or neutron star \citep{Dexter2013, Moriya2018}, a collapsar or jet \citep{MacFadyen1999}, or the spin-down energy of a rapidly-rotating magnetar \citep{Kasen2010, Woosley2010}.

In the magnetar model, the spin-down energy is emitted as a highly magnetized particle wind, which expands relativistically until it collides with the inner edge of the supernova ejecta, create forward and reverse shocks. The expanding wind becomes shocked by the reverse shock, accelerating the particles up to ultrarelativistic energies, which then emit via synchrotron radiation and inverse Compton scattering \citep{Gaensler2006}; this shocked wind is known as a pulsar wind nebula (PWN). The PWN applies a pressure to the supernova ejecta, causing it to accelerate, and the radiation from the nebula is thermalized in the ejecta, causing the ejecta temperature and supernova luminosity to both increase \citep{Kasen2010}. The PWN-ejecta interaction can also cause Rayleigh-Taylor instabilities, which can shred the inner ejecta and cause non-spherical structure to emerge in the ejecta \citep{Suzuki2017, Suzuki2021}.

The magnetar model predicts many multiwavelength signals that could be used to identify and characterize the newborn neutron star. Once the ejecta becomes optically thin, the non-thermal emission from the PWN can be detected directly, either at high energy in hard X-rays or gamma rays \citep{Kotera2013, Murase2015, Kashiyama2016}, or at low energy in radio \citep{Omand2018, Eftekhari2021, Murase2021}; two energetic SNe have radio detections at late times that are consistent with PWN emission, PTF10hgi \citep{Eftekhari2019, Law2019, Mondal2020, Hatsukade2021} and SN2012au \citep{Stroh2021}.  Dust formed in the supernova can absorb PWN emission and re-emit that energy in infrared \citep{Omand2019}, causing bright continuum emission recently seen in four SLSNe \citep{Chen2021, Sun2022}. High-energy photons and Rayleigh-Taylor induced mixing can change the chemical and ionization structure of the ejecta, leading to unique signatures in the supernova nebular spectrum \citep{Chevalier1992, Omand2023}. Aspherical ejecta caused by either hydro instabilities or an aspherical PWN can produce an optical polarization signal \citep{Tanaka2017}, which has been detected in some energetic SNe \citep{Inserra2016, Saito2020, Pursiainen2022, Poidevin2023, Pursiainen2023}, but not others \citep{Leloudas2015, Lee2019, Lee2020, Poidevin2022, Poidevin2023, Pursiainen2023}.

Accurate parameter estimation from the light curve around optical peak is essential for a number of reasons. Firstly, new surveys such as the LSST will detect SLSNe out to high redshift, where they can not all be classified with spectroscopy. Therefore, these supernovae will have to be characterized from their light curve data alone. Also, predicting late-time multiwavelength signals requires accurate parameter estimation, as different parameters that produce similar optical light curves can produce vastly different multiwavelength signals \citep[e.g.][]{Omand2018, Omand2019}. In most widely used parameter inference codes \citep[e.g][]{Nicholl2017}, the model used for inference of magnetar-driven SNe makes several assumptions to reduce computational complexity, which are unjustified outside a small region of the parameter space. In particular, they assume a constant ejecta velocity, which is independent of the magnetar luminosity, although numerical simulations show that ejecta acceleration due to PWN pressure plays a vital role in the dynamics of the ejecta \citep{Chen2016, Suzuki2017, Suzuki2019, Chen2020, Suzuki2021}. They also assume the magnetar spin down through pure vacuum dipole emission, even though studies of Galactic pulsars \citep{Lyne2015, Parthasarathy2020} and putative magnetars born in GRBs \citep{Lasky2017, Sarin2020b, Sarin2020a} show that most neutron stars are inconsistent with a pure vacuum dipole.

In this work, we present a model where these assumptions are relaxed, which can fully explore the diversity of magnetar-driven supernovae and unite phenomenologically different supernovae, such as SNe Ic-BL and SLSNe, under one theoretical framework. In Section \ref{sec:mod}, we introduce our model for magnetar-driven supernovae. In Section \ref{sec:result}, we show the diversity of supernovae and supernova observables resulting from differences in initial parameters. In Section \ref{sec:cases}, we perform Bayesian inference on a few varying supernovae to show how our model can consistently reproduce and explain them. Finally, in Section \ref{sec:conc}, we discuss the model's implications and conclude. Throughout the paper, we use the notation $Q_x = Q/10^x$ in cgs units unless otherwise noted.

\section{Model} \label{sec:mod}

The physics of our model is based on previous magnetar-driven kilonovae models \citep[e.g.][]{Yu2013, Metzger2019, Sarin2022}, but modified to describe supernovae.  We present here a non-relativistic model description, although the model implementation is fully relativistic.  For the fully relativistic description of the kinematics (Equations \ref{eqn:eej}, \ref{eqn:deintdt}, \ref{eqn:dvdt2}, \ref{eqn:lbol_pret}, \ref{eqn:lbol_postt}, \ref{eqn:opdep}, and \ref{eqn:tdif}), see \citet{Sarin2022}. We note that relativistic corrections are largely unimportant for supernovae apart from transients with exceptionally low ejecta masses and powerful magnetar engines. 
% Relativity will only be important when examining transients with very low ejecta masses, as the Lorentz factor of 0.1$M_\odot$ of ejecta with a kinetic energy of 10$^{53}$ erg is only $\gamma \approx 1.5$.

\subsection{Model Physics}

The central magnetar spins down by releasing its rotational energy 
\begin{equation}
    E_{\rm rot} = \frac{1}{2}I\Omega^2,
    \label{eqn:erot}
\end{equation}
where $I$ is the moment of inertia of the magnetar and $\Omega$ is the rotational angular frequency of the magnetar.  The time derivative of this relation gives the spin down luminosity, 
\begin{equation}
    L_{\rm SD} = I\Omega \dot{\Omega},
    \label{eqn:liood}
\end{equation}
which, given $\dot{\Omega} \propto -\Omega^n$ for braking index $n$, can be modelled generally as \citep{Lasky2017}
\begin{equation}
    L_{\rm SD}(t) = L_0 \left( 1 + \frac{t}{t_{\rm SD}} \right)^{\frac{1+n}{1-n}},
    \label{eqn:llasky}
\end{equation}
where $L_0$ is the initial magnetar spin-down luminosity and $t_{\rm SD}$ is the magnetar spin-down time.  The vacuum dipole spin-down mechanism \citep{Ostriker1969, Goldreich1969} has a corresponding braking index of $n = 3$, which gives a late-time spin-down down luminosity of $L \propto t^{-2}$ \citep{Zhang2001}, while gravitational wave spin down via magnetic deformation \citep{Cutler2000} has a corresponding braking index of $n = 5$, and has a late-time spin-down luminosity of $L \propto t^{-3/2}$.  There are also several other mechanisms for spin down, including particle winds \citep{Harding1999, Xu2001, Wu2003, Contopoulos2006}, multipolar radiation \citep{Petri2015}, and r-mode oscillations \citep{Ho2000}, among others \citep{Blandford1988, Lin2004, Ruderman2005, Chen2006, Yue2007, Contopoulos2007}, and several mechanisms can spin-down the magnetar simultaneously. We note that in general, $n$ is also expected to be variable during the early life of the magnetar \citep{Lander2018, Lander2020}, but we keep it constant for simplicity.  Integrating Equation \ref{eqn:llasky} gives the total rotational energy as a function of braking index:

\begin{equation}
    E_{\rm rot} = \frac{n-1}{2}L_0t_{\rm SD} . 
    \label{eqn:elt}
\end{equation}
This recovers $E_{\rm rot} = L_0t_{\rm SD}$ for vacuum dipole spin-down.

The rotational energy from the magnetar is converted into a pulsar wind.  This highly magnetized, ultrarelativistic wind collides with and pushes a shock into the supernova ejecta, increasing its kinetic and internal energy.  The internal energy is also increased by the absorption of PWN photons by the ejecta.  The total energy of the system is the combination of the kinetic and internal energy
\begin{equation}
    E_{\rm ej} = \frac{1}{2}M_{\rm ej} v_{\rm ej}^2 + E_{\rm int},
    \label{eqn:eej}
\end{equation}
where $M_{\rm ej}$ is the ejecta mass.  The evolution of this system is governed by the energy sources, radioactive heating and magnetar spin-down luminosity, and energy losses from radiated luminosity and adiabatic cooling from expansion.  The evolution of the internal energy of the ejecta is written as \citep{Kasen2016} 
\begin{equation}
    \frac{dE_{\rm int}}{dt} = \xi L_{\rm SD} + L_{\rm ra} - L_{\rm bol} - \mathcal{P}\frac{dV}{dt},
    \label{eqn:deintdt}
\end{equation}
where $L_{\rm ra}$ and $L_{\rm bol}$ are the radioactive power and emitted bolometric luminosity, respectively, $\xi$ is the fraction of spin-down luminosity injected into the ejecta, and $\mathcal{P}$ and $V$ are the pressure and volume of the ejecta.

Here, we adopt the \citet{Wang2015} prescription for gamma-ray leakage used in other models \citep[e.g.][]{Nicholl2017, Sarin2022} with 
\begin{equation}
   \xi = 1 - e^{-At^{-2}},
   \label{eq:xi}
\end{equation}
where
\begin{equation}
    A = \frac{3 \kappa_\gamma M_{\rm ej}}{4\pi v^2_{\rm ej}}
    \label{eq:leakage}
\end{equation}
is the leakage parameter and $\kappa_\gamma$ is the gamma-ray opacity of the ejecta.

The radioactive power from the decay of $^{56}$Ni is given by 
\begin{equation}
    L_{\rm ra} = f_{\rm Ni}M_{\rm ej}(L_{\rm ^{56}Ni} e^{-t/t_{\rm ^{56}Ni}} + L_{\rm ^{56}Co} e^{-t/t_{\rm ^{56}Co}}
    ),
    \label{eqn:lra}
\end{equation}
where $f_{\rm Ni}$ is the nickel fraction of the ejecta, $L_{\rm ^{56}Ni} = 6.45 \times 10^{43}$ erg s$^{-1}$ $M_\odot^{-1}$ and $L_{\rm ^{56}Co} = 1.45 \times 10^{43}$ erg s$^{-1}$ $M_\odot^{-1}$ are the decay luminosities of $^{56}$Ni and $^{56}$Co, and $t_{\rm ^{56}Ni} = 8.8$ days and $t_{\rm ^{56}Co} = 111.3$ days are the decay timescales for $^{56}$Ni and $^{56}$Co \citep{Nadyozhin1994}.  The inclusion of $L_{\rm ra}$ here allows our model to simplify to an \citet{Arnett1982} model for low $L_{\rm SD}$.

The dynamical evolution of the ejecta is given by \citep{Sarin2022}
\begin{equation}
    \frac{dv_{\rm ej}}{dt} = \frac{c^2\mathcal{P}(d\mathcal{V}/dt)}{M_{\rm ej}v_{\rm ej}^3},
    \label{eqn:dvdt}
\end{equation}
where
\begin{align}
    \mathcal{V} &= \frac{4}{3}\pi R_{\rm ej}^3, \\
    \frac{d\mathcal{V}}{dt} &= 4\pi R_{\rm ej}^2 v_{\rm ej}, \\
    \mathcal{P} &= \frac{E_{\rm int}}{3\mathcal{V}}.
\end{align}
Substituting these into Equation \ref{eqn:dvdt} gives
\begin{equation}
    \frac{dv_{\rm ej}}{dt} = \frac{c^2 E_{\rm int}}{M_{\rm ej}R_{\rm ej}v_{\rm ej}^2}.
    \label{eqn:dvdt2}
\end{equation}
The initial ejecta velocity is set by
\begin{equation}
    v_{\rm ej,0} = \sqrt{\frac{2E_{\rm SN}}{M_{\rm ej}}},
\end{equation}
where $E_{\rm SN}$ is the supernova explosion energy.

The bolometric radiated luminosity is \citep{Kasen2010, Kotera2013}
\begin{align}
    L_{\rm bol} = & \frac{E_{\rm int}c}{\tau R_{\rm ej}} = \frac{E_{\rm int}t}{t_{\rm dif}^2} & (t \leq t_\tau), \label{eqn:lbol_pret} \\
    = & \frac{E_{\rm int}c}{R_{\rm ej}}, & (t > t_\tau),\label{eqn:lbol_postt}
\end{align}
where 
 \begin{equation}
     \tau = \frac{\kappa M_{\rm ej} R_{\rm ej}}{\mathcal{V}}
     \label{eqn:opdep}
 \end{equation}
is the optical depth of the ejecta, $\kappa$ is the ejecta opacity, 
\begin{equation}
    t_{\rm dif} = \left(\frac{\tau R_{\rm ej} t}{c}\right)^{1/2}
    \label{eqn:tdif}
\end{equation}
is the effective diffusion time, and $t_\tau > t_{\rm dif}$ is the time when $\tau = 1$.

Calculating the bolometric luminosity of the magnetar-driven transient (Equations \ref{eqn:lbol_pret} and \ref{eqn:lbol_postt}) involves solving the evolution of the internal energy and dynamics (Equations \ref{eqn:deintdt} and \ref{eqn:dvdt2} respectively) using the input power sources (Equations \ref{eqn:llasky} and \ref{eqn:lra}).  The photospheric temperature is determined from the bolometric luminosity and ejecta radius until the temperature reaches the photospheric plateau temperature, as in \citet{Nicholl2017}.  This can be expressed as 

\begin{equation}
T_{\rm phot} (t) = 
\begin{cases}
    \left( \frac{L_{\rm bol}(t)}{4 \pi \sigma R_{\rm ej}^2}\right)^{1/4} & \text{ for } \left( \frac{L_{\rm bol}(t)}{4 \pi \sigma R_{\rm ej}^2}\right)^{1/4} > T_{\rm min}, \\
    T_{\rm min} & \text{ for } \left( \frac{L_{\rm bol}(t)}{4 \pi \sigma R_{\rm ej}^2}\right)^{1/4} \leq T_{\rm min}
\end{cases}
\label{eqn:tphot}
\end{equation}

The spectral energy distribution (SED) of the transient is then calculated using the cutoff blackbody used in \citet{Nicholl2017}, with $F_{\lambda < \lambda_{\rm cut}} = F_\lambda (\lambda/\lambda_{\rm cut})$ and $\lambda_{\rm cut} = 3000 \AA$ \citep{Chomiuk2011, Nicholl2017b}, but this can also be switched to a simple blackbody.

\subsection{Parameters, Priors, and Implementation}

The model presented above is implemented into the open-source electromagnetic transient fitting software package, {\sc{Redback}} \citep{sarin23_redback}.  The input parameters for the model, their default priors, and the values used in Section \ref{sec:result} are listed in Table \ref{tbl:modparam}.

\begin{table*}
\centering
\begin{tabular}{ccccc}
   Parameter & Definition & Units & Default Prior & Section \ref{sec:etos} Value \\ \hline
   $L_0$ & Initial Magnetar Spin-Down Luminosity & erg s$^{-1}$ & L[$10^{40}$, $10^{50}$] & Varying \\
   $t_{\rm SD}$ & Spin-Down Time & s & L[$10^{2}$, $10^{8}$] & Varying \\
   $n$ & Magnetar Braking Index &  & U[1.5, 10] & 3 \\
   $f_{\rm Ni}$ & Ejecta Nickel Mass Fraction &  & L[10$^{-3}$, 1] & 0 \\
   $M_{\rm ej}$ & Ejecta Mass & $M_{\odot}$ & L[10$^{-2}$, 1 $\times$ 10$^2$] & Varying \\
   $E_{\rm SN}$ & Supernova Explosion Energy & erg & U[$5 \times 10^{50}$, $2 \times 10^{51}$] & $10^{51}$ \\
   $\kappa$ & Ejecta Opacity & cm$^2$ g$^{-1}$ & U[0.05, 0.2] & 0.1 \\
   $\kappa_\gamma$ & Ejecta Gamma-Ray Opacity & cm$^2$ g$^{-1}$ & L[$10^{-4}$, $10^{4}$] & 0.1 \\
   $T_{\rm min}$ & Photospheric Plateau Temperature & K & U[$3 \times 10^{3}$,  $10^{4}$] & 5000 \\
\end{tabular}
\caption{The parameters and default priors for the generalized magnetar model, as well as the values used for the parameter exploration in Section \ref{sec:result}. Priors are either uniform (U) or log-uniform (L).}
\label{tbl:modparam}
\end{table*}

The range of the default priors can lead to magnetars with rotational energies greater than 10$^{58}$ erg, which is unphysically high.  When performing inference, the prior can be restricted to only sample parameters where $E_{\rm rot}$ is below a certain threshold.  We recommend using this constraint to prevent unphysically high energies, and use a value of 10$^{53}$ erg for the case studies presented in Section \ref{sec:cases}.

Due to the acceleration of the ejecta from the PWN, $v_{\rm ej}$ is not constant and can not be used as a free parameter, since it is coupled to both $E_{\rm SN}$ and $L_{\rm SD}$.  Velocity information from spectroscopy can be used to weight the results, although the velocity measured from absorption widths is not the same as the photospheric velocity which is not the same as the ejecta velocity \citep{Arnett1982}, so we caution against this unless the velocities are well calibrated \citep[e.g.][]{Dessart2016}.  
%If one does want to use a velocity prior during inference, the model likelihoods can be reweighted as a post-processing step to account for a prior on the photospheric velocity at peak.

Our model differs from others \citep[e.g.][]{Nicholl2017} in the choice of input parameters used to determine the magnetar luminosity.  Previous models use the initial magnetar spin period $P_0$, dipole component of the pulsar magnetic field $B$, and neutron star mass $M_{\rm NS}$; while we use $L_0$ and $t_{\rm SD}$ (and $n$, which is implicitly fixed to 3 in other models).  The interpretation of braking index can be complicated due to the number of possible spin-down mechanisms, so users wanting to test specific mechanisms can swap different magnetar implementations due to the modular nature of the model and {\sc{Redback}}.
We note that as the magnetar luminosity for vacuum dipole spin-down can determined from only $L_0$ and $t_{\rm SD}$ (see Equation \ref{eqn:llasky}), using three parameters is unnecessary for parameter inference. Using these parameters also avoids assumptions such as the moment of inertia of the neutron star, which depends on the equation of state (EoS) \citep{Lattimer2005} and can vary depending on the mass and spin period of the neutron star \citep{Worley2008}.  To recover parameters such as the magnetar spin period and magnetic field, one can use the scalings used in other models such as 

\begin{align}
    L_0 = & 2.0 \times 10^{47} P_{0, -3}^{-4} B_{14}^2 \label{eqn:l0scale}, \\
    t_{\rm SD}= & 1.3 \times 10^5 P_{0, -3}^2 B_{14}^{-2} \left( \frac{M_{\rm NS}}{1.4 M_\odot} \right) \label{eqn:tsdscale}
\end{align}
for transients consistent with $n = 3$, which assumes a neutron star with moment of inertia of $\sim$ 1.3 $\times$ 10$^{45}$ g cm$^2$ for a 1.4 $M_\odot$ neutron star, which is consistent with the APR EoS \citep{Akmal1998} or MDI EoS with density independent nuclear symmetry energy \citep{Das2003, Shetty2007}, which both give neutron star radii $\sim$ $11.5-12$ km \citep{Worley2008}.  These scalings become more complicated for $n \neq 3$ \citep[e.g.,][]{Shapiro1983}, with other dependencies such as the ellipticity of the neutron star or bulk viscosity (depending on the spin-down processes involved), leaving it difficult to definitively recover a spin period or magnetic field with such simplified models.

\section{Results} \label{sec:result}

We now explore the diversity of magnetar driven-supernovae for different initial conditions using the model derived in Section \ref{sec:mod}.  We assume the supernova explosion energy is 10$^{51}$ erg, typical for a neutrino-driven explosion, and the supernova is entirely powered by magnetar energy after the initial explosion, with no contribution from $^{56}$Ni.  We fix both the ejecta optical opacity and gamma-ray opacity to 0.1 cm$^2$ g$^{-1}$. This optical opacity value is typical of stripped-envelope supernovae \citep{Inserra2013, Kleiser2014} and the gamma-ray opacity will not affect the peak light curve properties unless it is extremely low, so we use the same value for both opacities here for simplicity.  We also fix the photospheric plateau temperature to 5000 K, which is typical for SLSNe \citep{Nicholl2017} and SNe Ic-BL \citep{Taddia2019}, although slightly lower than the temperature of 6000 K suggested by \citet{Nicholl2017}, which corresponds to the recombination temperature of O II; ultimately, this temperature discrepancy will have no impact on the peak light-curve properties in most cases, and will only affect the light curve at late times.

\begin{figure}
\centering
\includegraphics[width=0.8\linewidth]{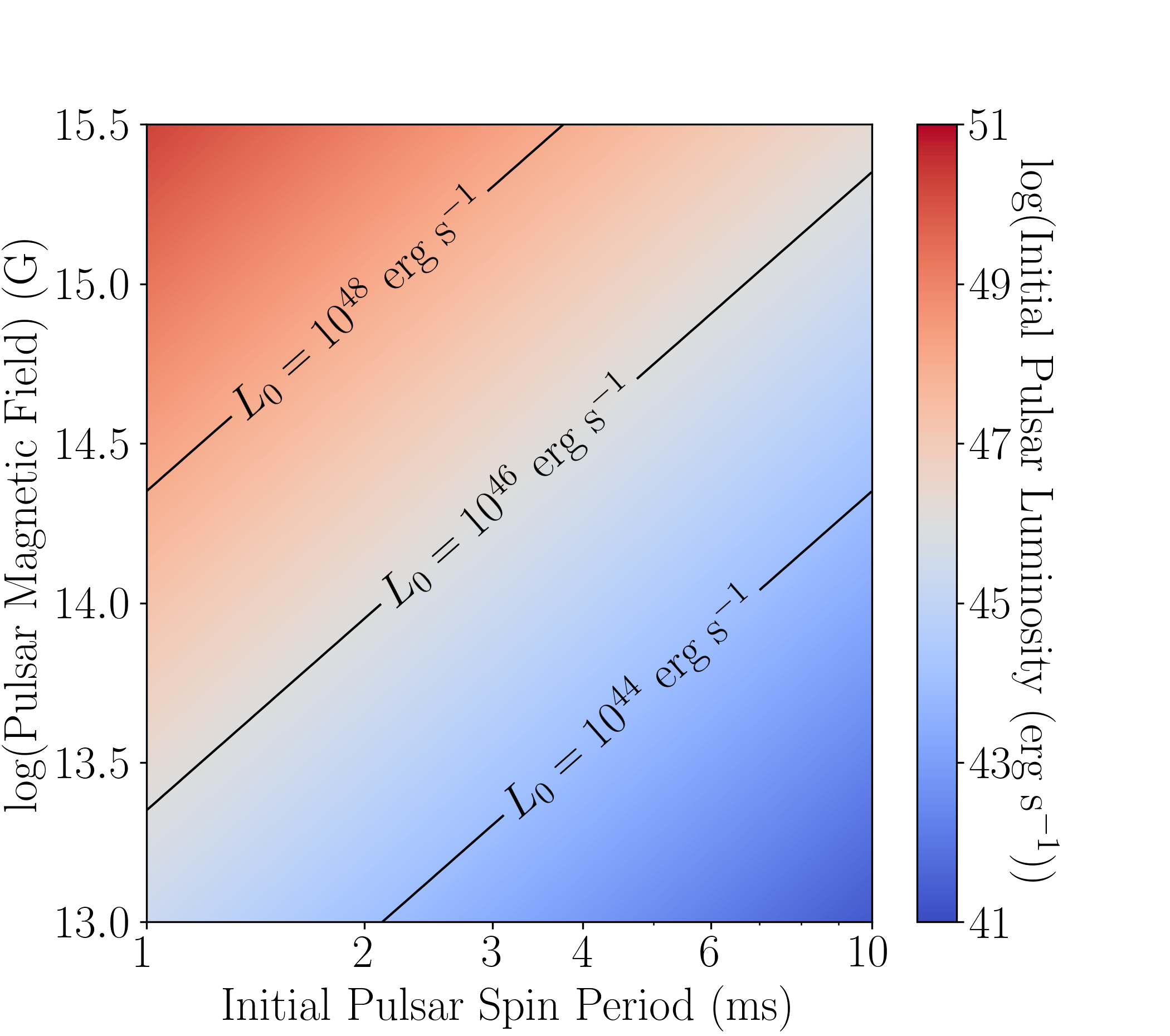} \\
\includegraphics[width=0.8\linewidth]{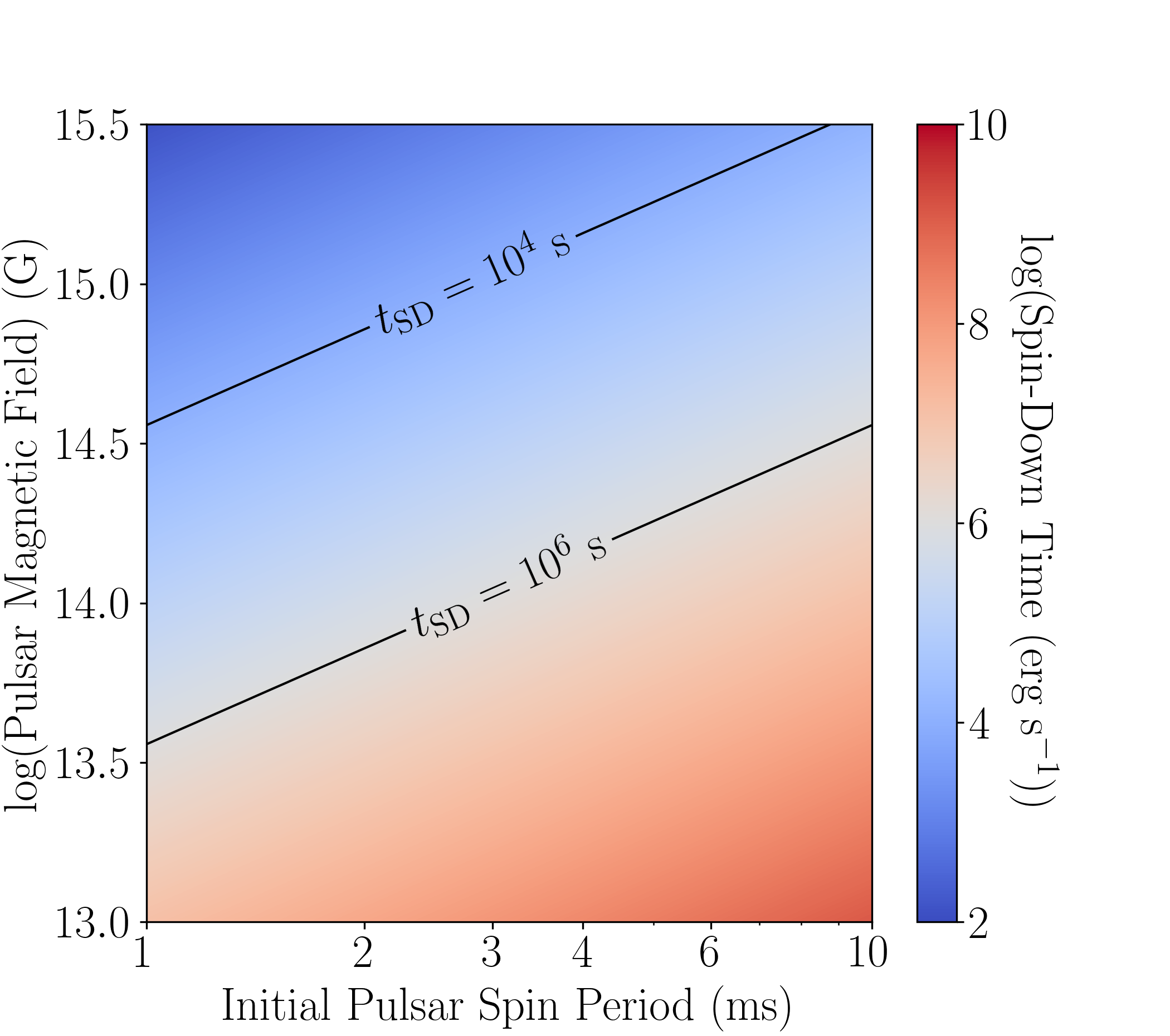}
\caption{Initial pulsar luminosities $L_0$ (top) and spin-down times $t_{\rm SD}$ (bottom) for different initial pulsar rotation periods and magnetic fields.}%
\label{fig:parammap}
\end{figure}

First, we show the diversity of supernovae that can be produced using this model in Section \ref{sec:etos}.  To compare with previously derived results, we assume vacuum dipole spin-down and use Equations \ref{eqn:l0scale} and \ref{eqn:tsdscale} with a 1.4 $M_\odot$ neutron star to express our initial conditions in terms of $P_0$ and $B$; the mapping between ($L_0$, $t_{\rm SD}$) and ($P_0$, $B$) is shown in Figure \ref{fig:parammap}.  Then we show the effect of changing braking index in Section \ref{sec:brake}.

%\centering%

\subsection{Energetics, Timescales, and Observables} \label{sec:etos} 

We first see if our model can reproduce typical observables for the observed populations of SLSNe and SNe Ic-BL.  SLSNe at optical light curve peak typically have g-band absolute magnitudes $-23 < M_g < -20$, with most brighter than $-$21; spectroscopically determined photospheric velocities of $\sim$ 10 000 $-$ 15 000 km s$^{-1}$; rise times of 10 $-$ 70 days; and $g - r$ of $-$0.3 $-$ 0.3 \citep{Chen2023a}.  SNe Ic-BL at optical light curve peak typically have r-band absolute magnitudes around $-20 < M_r < -18$, with most fainter than $-$19; spectroscopically determined photospheric velocities of $\sim$ 15 000 $-$ 30 000 km s$^{-1}$ \citep{Modjaz2016}; rise times of 5 $-$ 20 days; and $g - r$ of 0 $-$ 0.5 \citep{Taddia2019}.  The inferred ejecta masses of typical SLSNe and SNe Ic-BL are both around 5 $M_\odot$ \citep{Taddia2019, Chen2023b}, and SLSNe show a negative correlation between initial spin period and ejecta mass \citep{Blanchard2020}.  We also compare our results to the two dimensional radiation-hydrodynamic simulations of \citet{Suzuki2021}, who presented models with $L_0 = 10^{46}$ erg s${-1}$ (L46 models) and $L_0 = 10^{48}$ erg s${-1}$ (L48 models) and total magnetar rotational energies of 1, 3, and 10 $\times$ 10$^{51}$ erg.

Several inefficiencies can prevent all the magnetar spin-down luminosity from eventually being emitted as supernova luminosity.  Some fraction ($\xi$ from Equation \ref{eqn:deintdt}) will escape without interacting with the ejecta at all, and possibly be detected as non-thermal x-rays or gamma rays. 
Some fraction of the energy will also accelerate the ejecta instead of thermalizing and being re-emitted, which will affect both the supernova luminosity and peak timescale; this fraction can be determined by the ratio of the spin-down time and the diffusion time \citep{Suzuki2021, Sarin2022}.

\begin{figure*}
%\centering
%\settoheight{\tempdima}{\includegraphics[width=1.1\linewidth]{O6y_figs/OIfrac_T1e+05.png}}%
\newcolumntype{D}{>{\centering\arraybackslash} m{5cm}}
\noindent
\makebox[\textwidth]{
\begin{tabular}{DDD}
 \boldsymbol{$P_{\rm 0}$} \textbf{ = 1 ms} & \boldsymbol{$P_{\rm 0}$} \textbf{ = 3 ms} & \boldsymbol{$M_{\rm ej}$} \textbf{ = 10} \boldsymbol{$M_\odot$}\\
\includegraphics[width=1.1\linewidth]{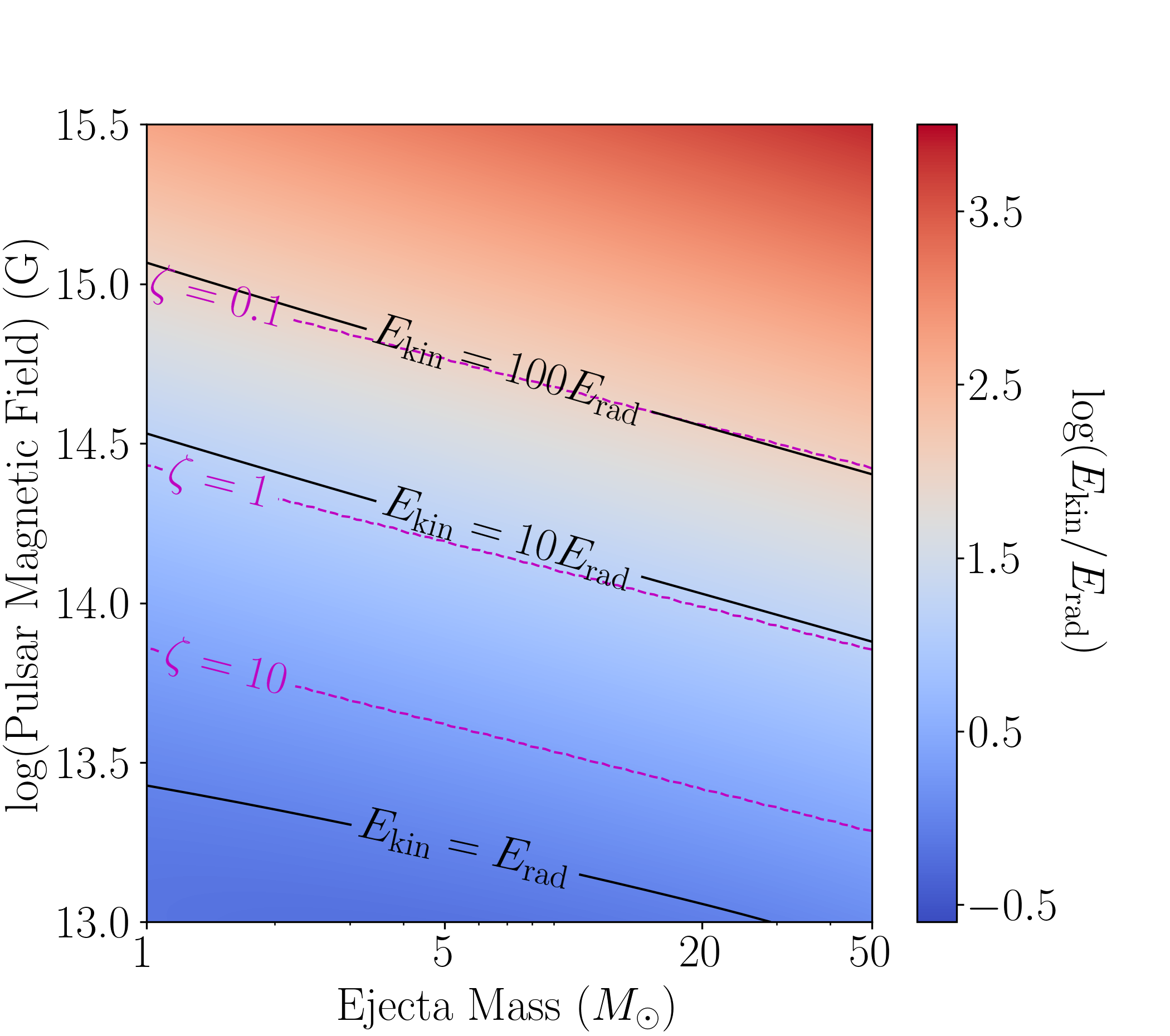}&
\includegraphics[width=1.1\linewidth]{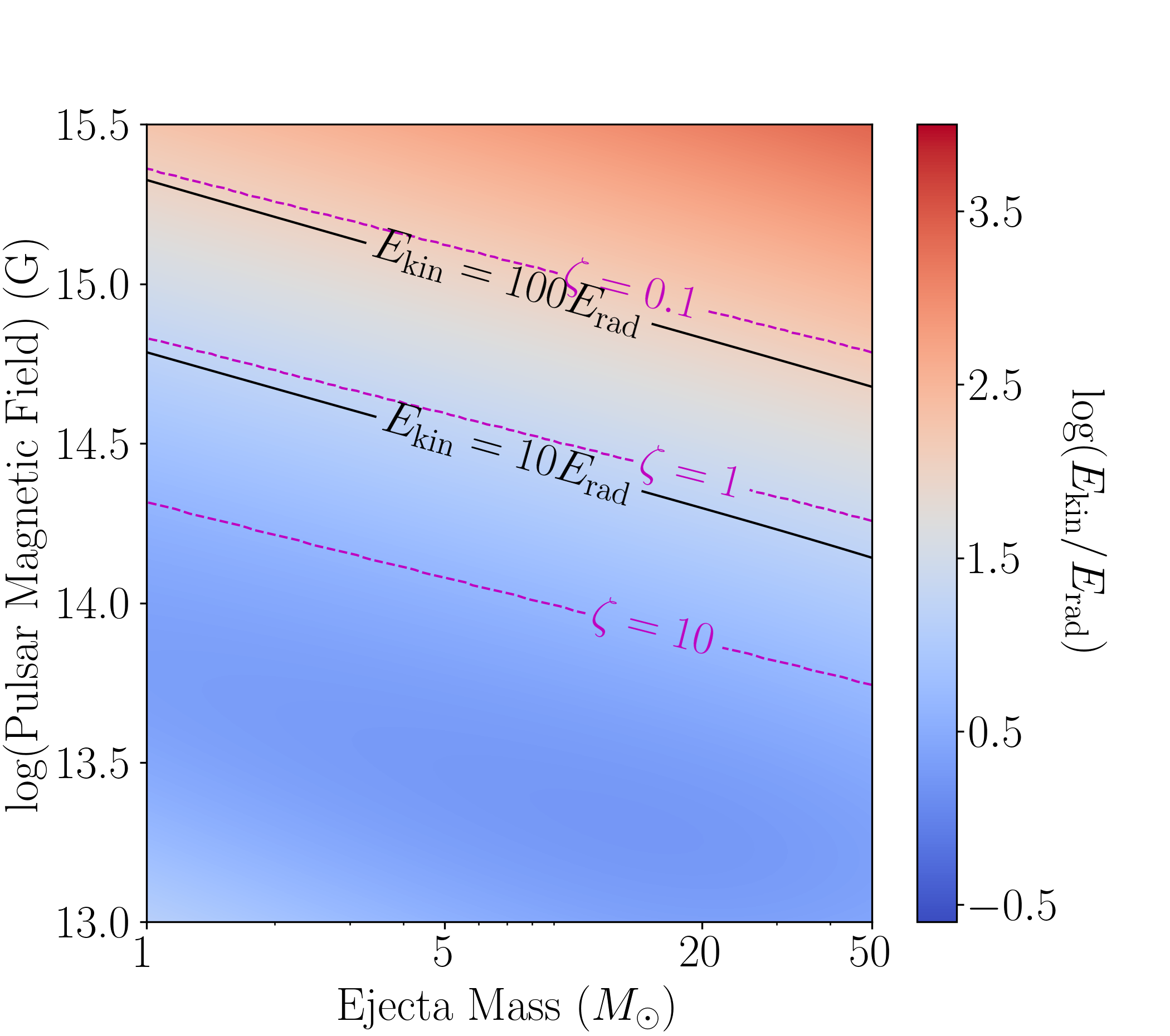}&
\includegraphics[width=1.1\linewidth]{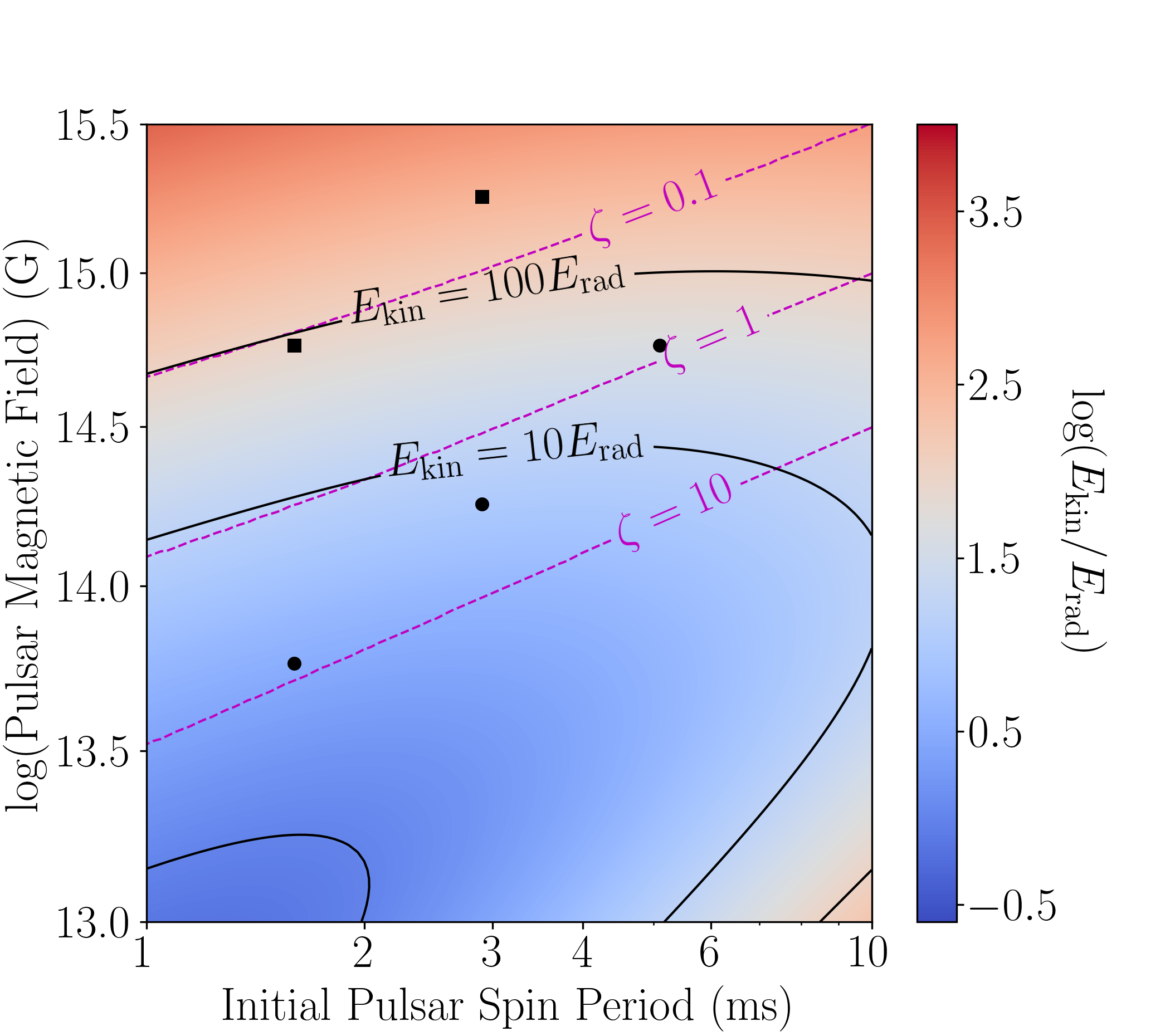}\\[-1.5ex]
\end{tabular}}
\caption{Ratio of kinetic to radiated energy for supernovae with varying ejecta mass and $P_0 = 1$ ms (left) and $P_0 = 3$ ms (middle) and with varying spin period and $M_{\rm ej} = 10 M_\odot$ (right).  The black lines indicate contours of constant $E_{\rm kin}/E_{\rm rad}$ (1,10, and 100), while the magenta lines indicate contours of constant $\zeta = t_{\rm SD}/t_{\rm dif}$ (0.1, 1, and 10).  The black symbols indicate the parameters used in \citet{Suzuki2021}, with the circles representing the L46 models and the squares representing the L48 models.}%
\label{fig:erat}
\end{figure*}

Figure \ref{fig:erat} shows the ratio of the final supernova kinetic and radiated energies for various magnetic fields and ejecta masses for spin periods of 1 ms (close to the mass shedding limit for neutron stars \citep{Watts2016}) and 3 ms (where the spin-down luminosity and explosion energy can become comparable), and for various magnetic fields and spin periods for an ejecta mass of 10 $M_\odot$, the same ejecta mass as \citet{Suzuki2021}.  The energy ratio $E_{\rm kin}$/$E_{\rm rad}$ does correlate strongly with $\zeta = t_{\rm SD}/t_{\rm dif}$ up to large ejecta mass for low spin periods, although as spin period increases, the correlation gets weaker as the behaviour of the energy ratio changes; this is due to the total amount of energy injected by the magnetar prior to the diffusion time decreasing below the explosion energy, meaning that the ejecta dynamics are no longer primarily determined by the magnetar spin-down energy.  There is a small region of the parameter space at low spin period, magnetic field, and ejecta mass where the radiated energy can surpass the kinetic energy. However, for most of the parameter space this ratio is between 1 and 100.  Typical supernovae have $E_{\rm kin}$/$E_{\rm rad}$ = 10$^{51}$ erg / 10$^{49}$ erg = 100, but without a contribution from $^{56}$Ni this ratio can go much higher.  \citet{Suzuki2021} find $E_{\rm kin}$/$E_{\rm rad}$ to be lower than we do in their 2D simulations by a factor of $\sim$ 2-10 depending on the model. However, they also find that multi-dimensional effects cause the supernova to be more luminous compared to 1D models at early times due to hot bubble breakout in the ejecta.

%\centering%

\begin{figure*}
%\centering
%\settoheight{\tempdima}{\includegraphics[width=1.1\linewidth]{O6y_figs/OIfrac_T1e+05.png}}%
\newcolumntype{D}{>{\centering\arraybackslash} m{5cm}}
\noindent
\makebox[\textwidth]{
\begin{tabular}{m{1cm} DDD}
& \boldsymbol{$P_{\rm 0}$} \textbf{ = 1 ms} & \boldsymbol{$P_{\rm 0}$} \textbf{ = 3 ms} & \boldsymbol{$M_{\rm ej}$} \textbf{ = 10} \boldsymbol{$M_\odot$}\\
\boldsymbol{$t_{\rm peak}$}&
\includegraphics[width=1.1\linewidth]{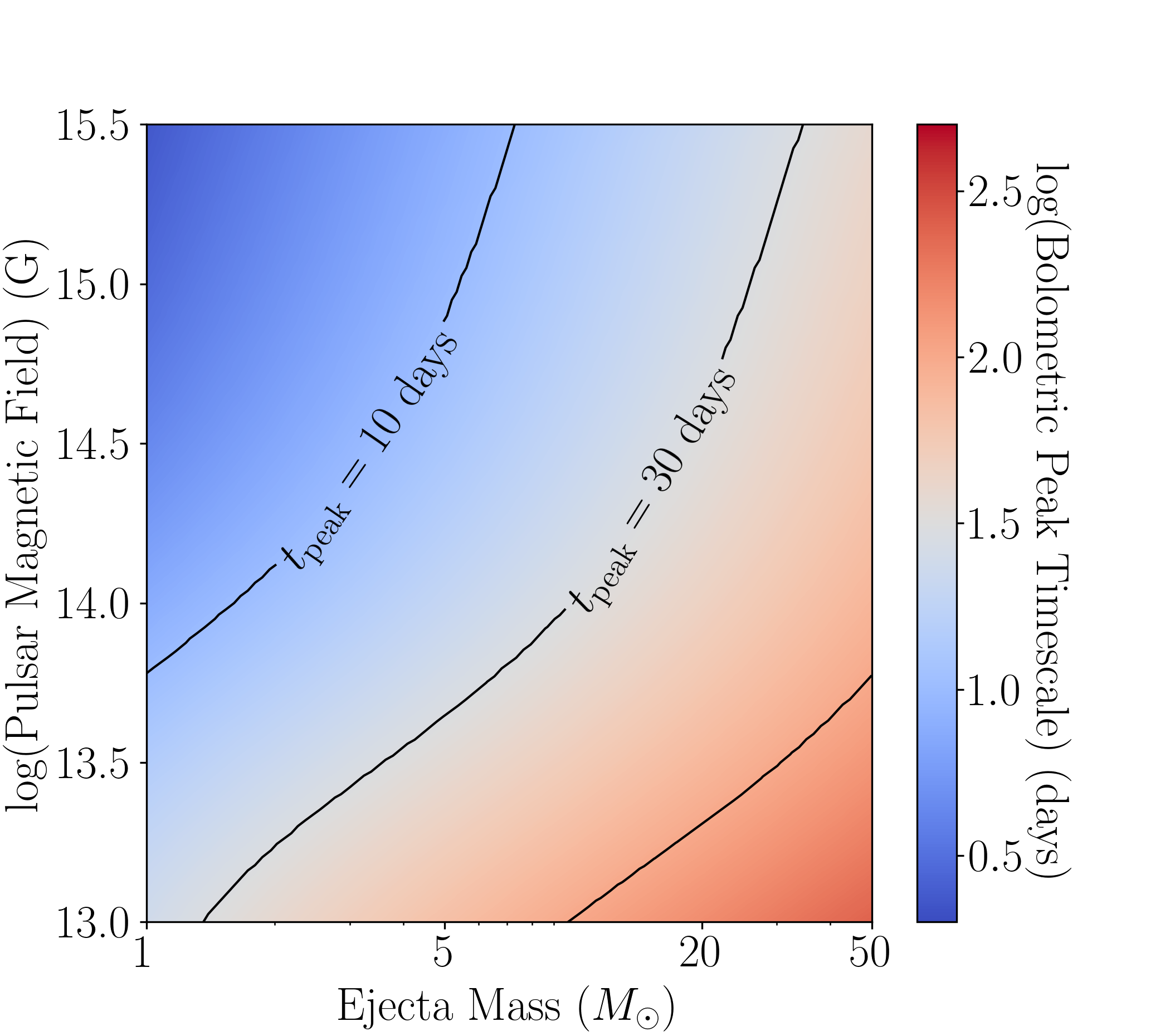}&
\includegraphics[width=1.1\linewidth]{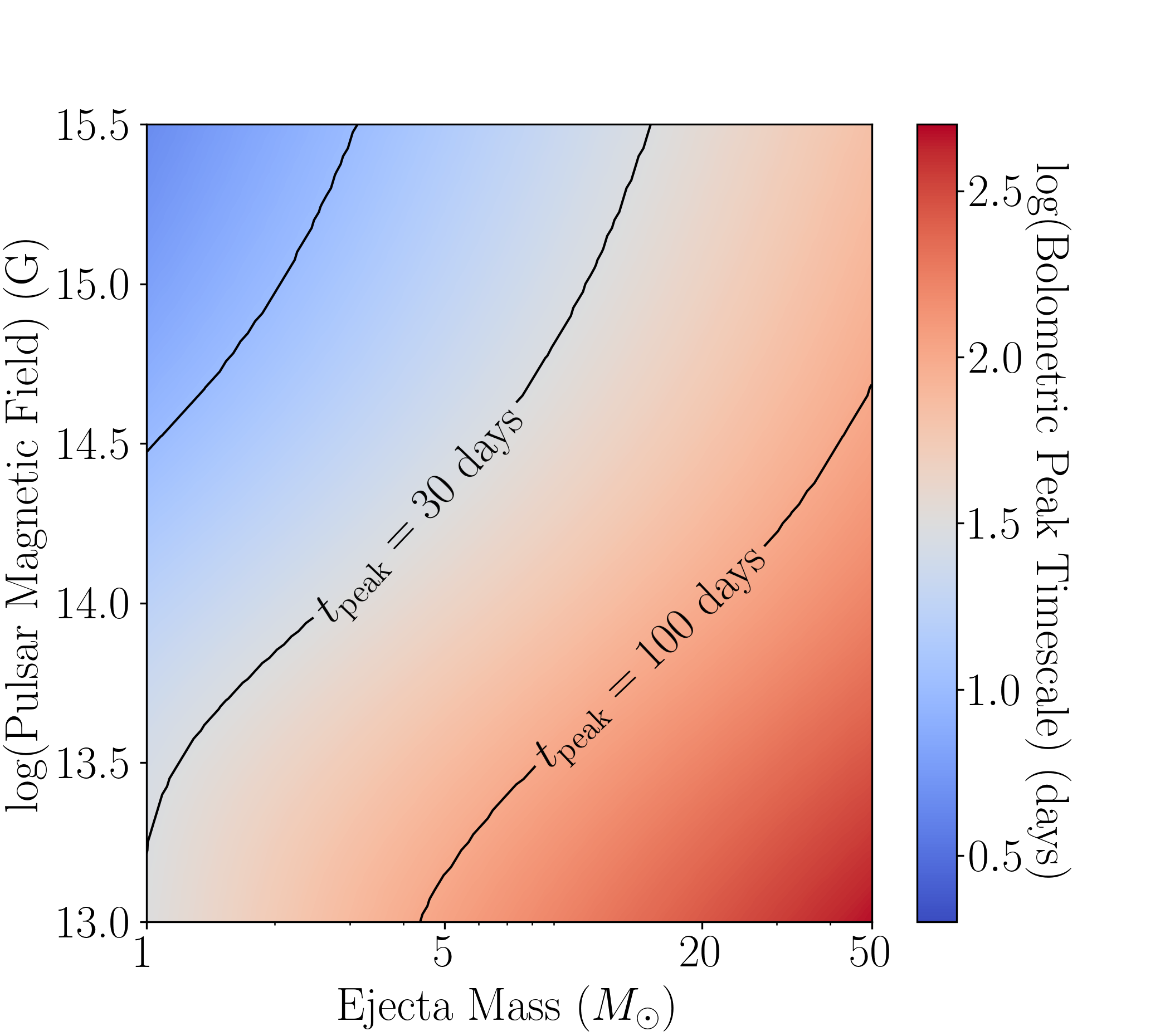}&
\includegraphics[width=1.1\linewidth]{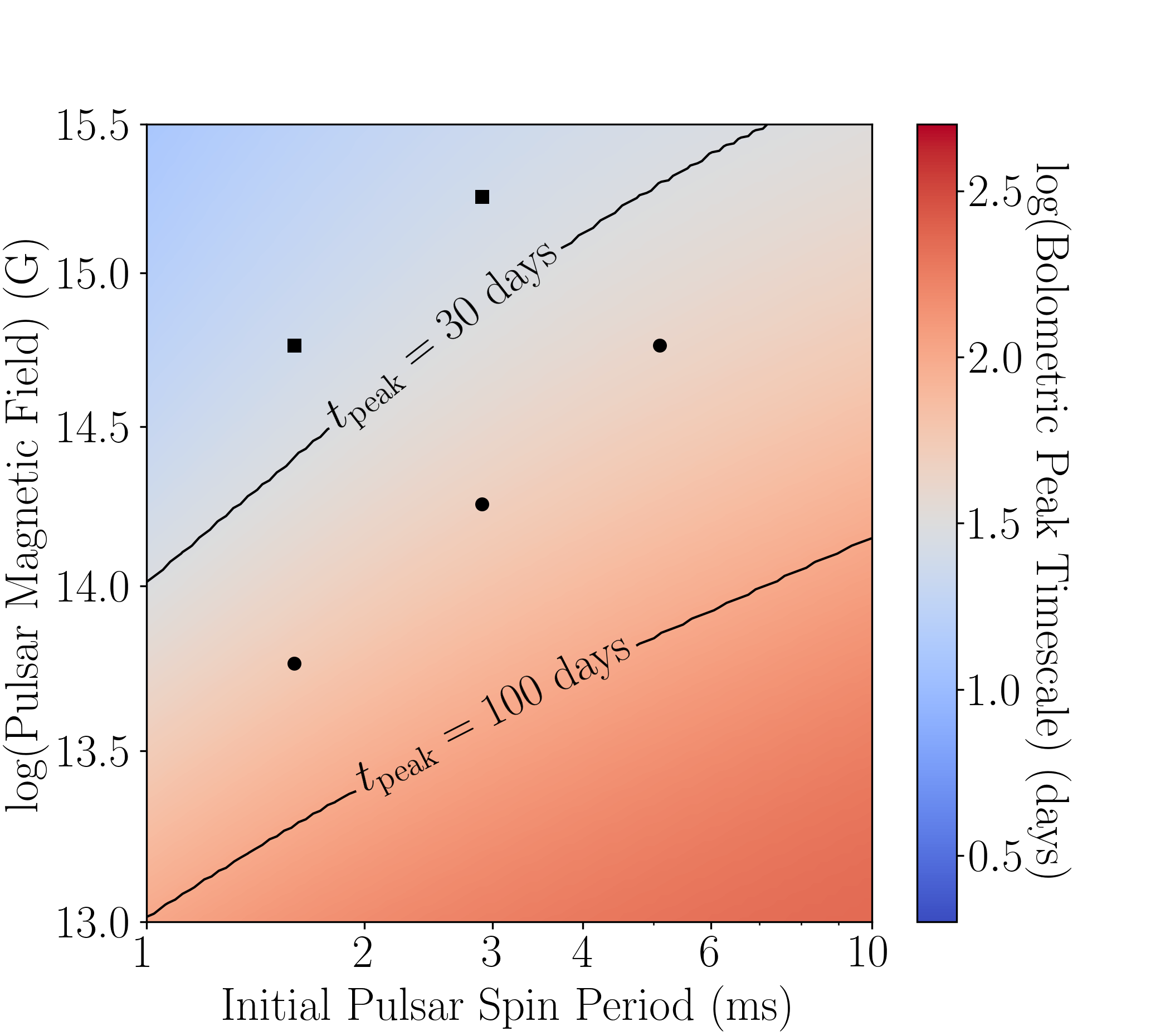}\\[-1.0ex]
\boldsymbol{$v_{\rm ej}$}&
\includegraphics[width=1.1\linewidth]{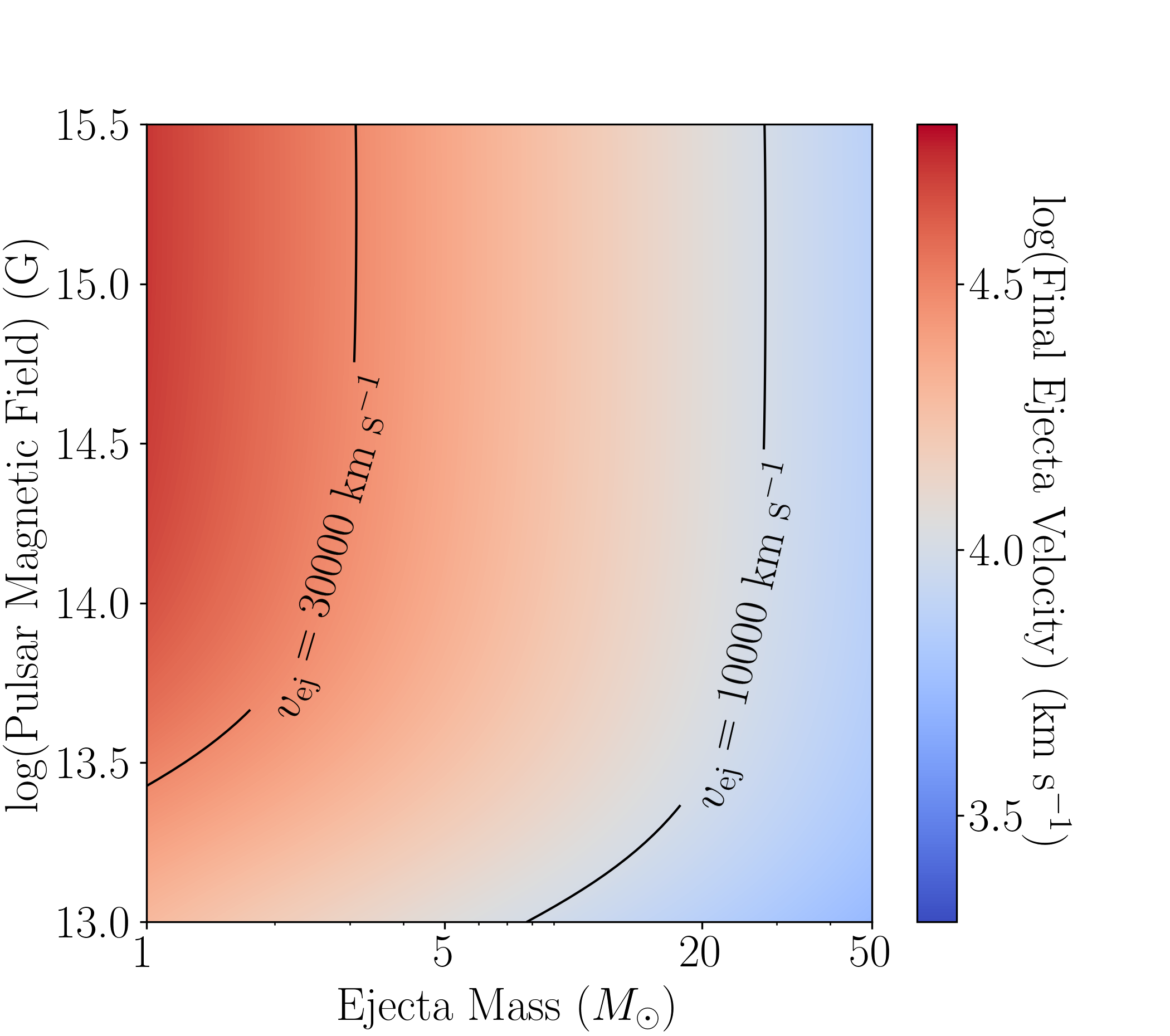}&
\includegraphics[width=1.1\linewidth]{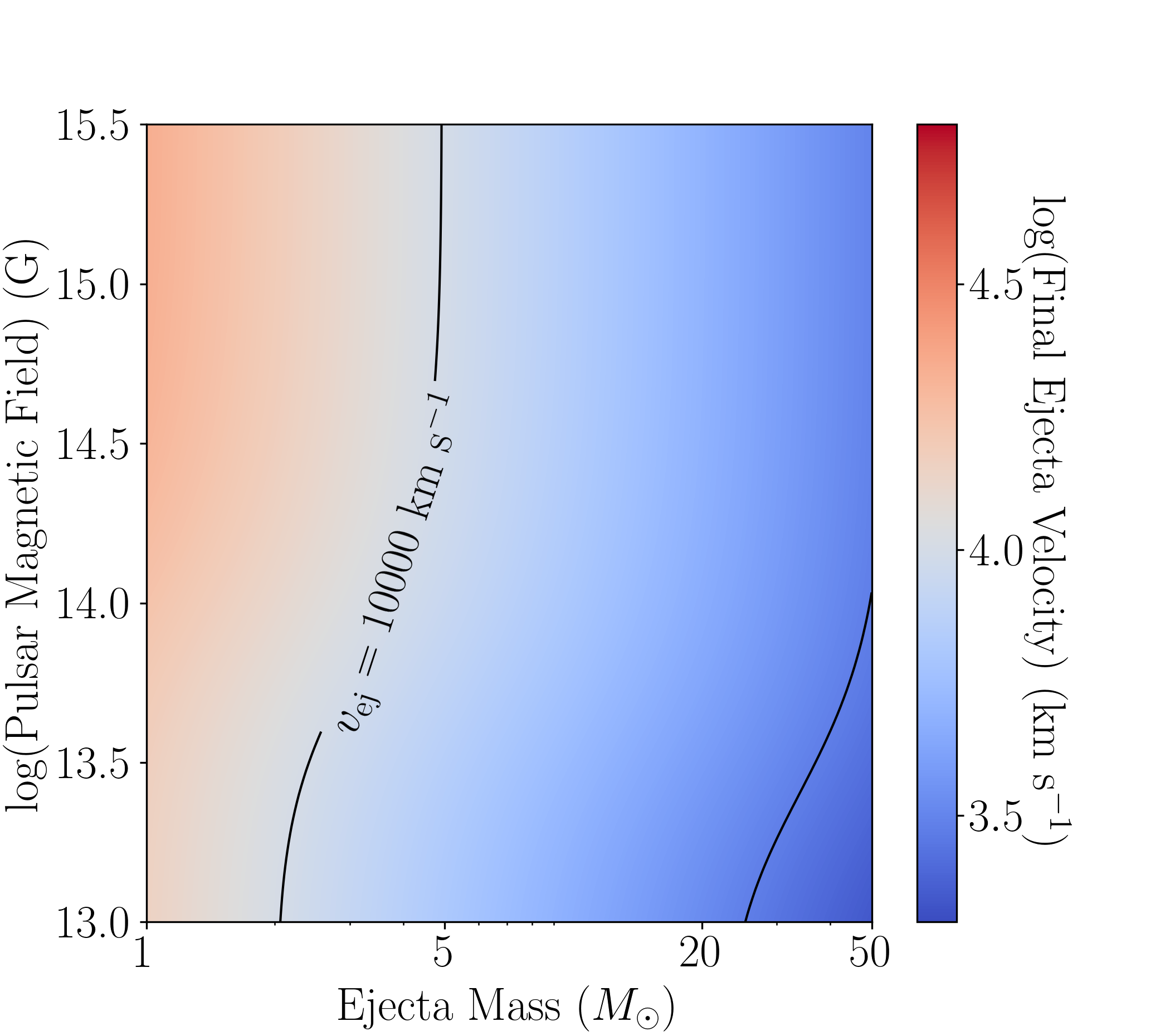}&
\includegraphics[width=1.1\linewidth]{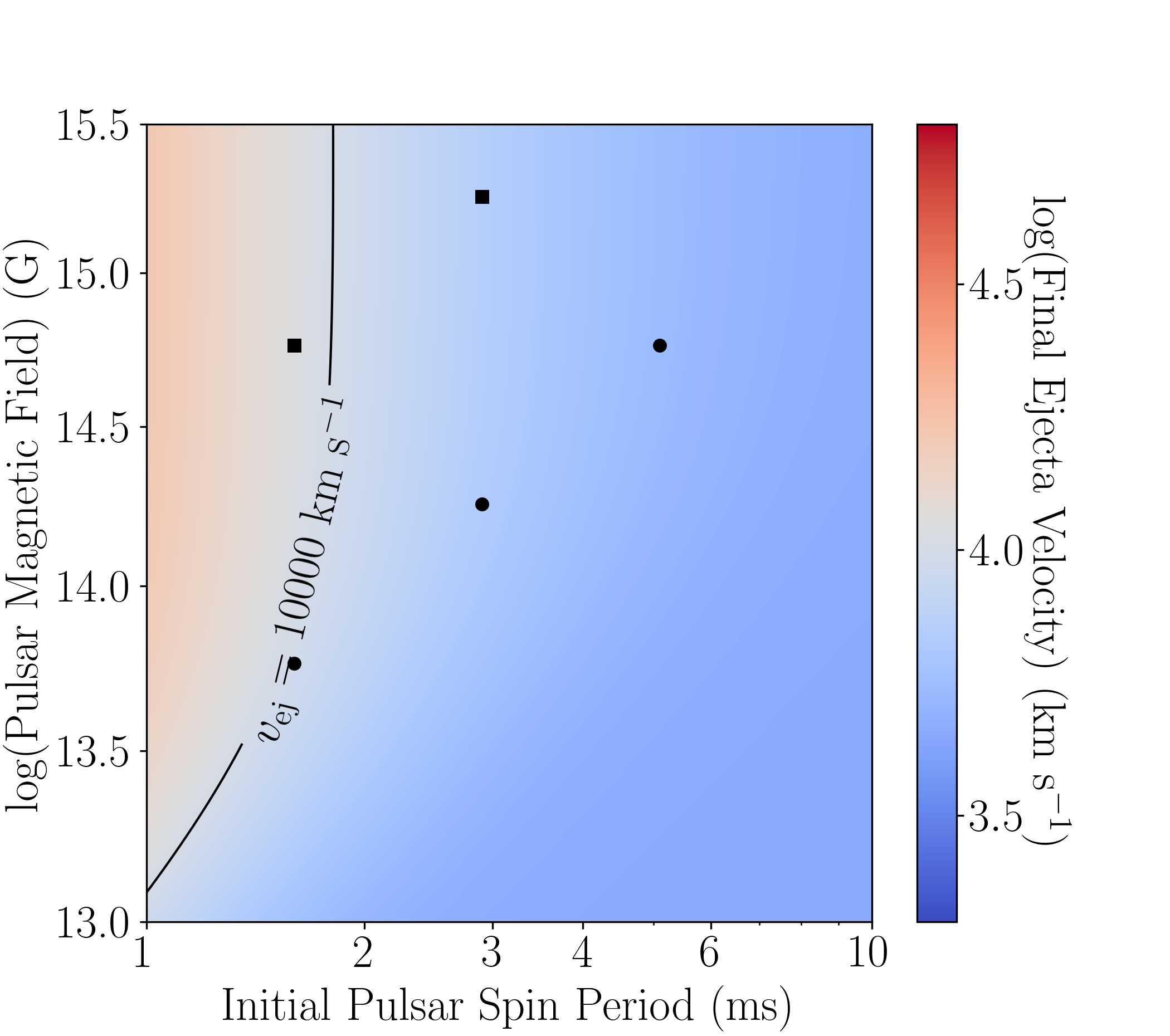}\\[-1.5ex]
\end{tabular}}
\caption{Bolometric peak timescale $t_{\rm peak}$ (top) and final ejecta velocity $v_{\rm ej}$ (bottom) for supernovae with varying ejecta mass and $P_0 = 1$ ms (left) and $P_0 = 3$ ms (middle) and with varying spin period and $M_{\rm ej} = 10 M_\odot$ (right).  The black lines indicate notable values of $t_{\rm peak}$ (10 ,30, and 100 days) and $v_{\rm ej}$ (3000, 10000, and 30000 km s$^{-1}$).  The black symbols indicate the parameters used in \citet{Suzuki2021}, with the circles representing the L46 models and the squares representing the L48 models.}%
\label{fig:timevej}
\end{figure*}

Figure \ref{fig:timevej} shows the peak timescale of the bolometric luminosity and the final ejecta velocity over the same parameter grid as Figure \ref{fig:erat}. The trends match our theoretical expectations, as $P_0$ increases, the total energy of the system decreases, causing the ejecta velocity to decrease and the peak timescale to increase.  Conversely, as $B$ increases, the spin-down time decreases, causing the ejecta velocity to increase and the peak timescale to decrease.  
Finally, as $M_{\rm ej}$ increases, the diffusion time increases, causing the ejecta velocity to decrease and the peak timescale to increase.
The final ejecta velocity is largely insensitive to the magnetic field strength above a certain field threshold despite a change in peak timescale. This is a product of the coupling of dynamical evolution of the ejecta to the magnetar's rotational energy, as the final velocity is reached earlier in the supernova evolution for higher magnetic field (due to their shorter spin-down timescale). 
For a magnetar with rotation period of 1 ms, a magnetic field of $>$ 10$^{15}$ is needed to reduce the peak timescale below 10 days, even at an ejecta mass of 1 $M_\odot$, meaning that the fastest SNe Ic-BL likely require both an ejecta mass below 1 $M_\odot$ and a magnetar spinning at close to breakup speeds.  
Timescales of $>$ 20 days require ejecta masses below 5 $M_\odot$, meaning that our model will likely estimate a lower ejecta mass for SNe Ic-BL than \citet{Taddia2019}.
Timescales and ejecta velocities typical of SLSNe can be reproduced over a large portion of the parameter space. 
However, a higher ejecta mass is required for faster spinning magnetars to keep the velocities below that of SNe Ic-BL. 
In contrast, a low ejecta mass is required for faster-spinning magnetars to keep the timescales below $\sim$ 100 days and the velocities higher than $\sim$ 10 000 km s$^{-1}$, providing some phenomenological justification for the mass-spin correlation found by \citet{Blanchard2020}.
\citet{Suzuki2021} find faster timescales compared to us in the L46 2D simulations, but similar timescales in their 1D simulations, and an average ejecta velocity that is roughly consistent with our model.

%\centering%

\begin{figure*}
%\centering
%\settoheight{\tempdima}{\includegraphics[width=1.1\linewidth]{O6y_figs/OIfrac_T1e+05.png}}%
\newcolumntype{D}{>{\centering\arraybackslash} m{5cm}}
\noindent
\makebox[\textwidth]{
\begin{tabular}{m{1cm} DDD}
& \boldsymbol{$P_{\rm 0}$} \textbf{ = 1 ms} & \boldsymbol{$P_{\rm 0}$} \textbf{ = 3 ms} & \boldsymbol{$M_{\rm ej}$} \textbf{ = 10} \boldsymbol{$M_\odot$}\\
\boldsymbol{$M_{\rm g}$}&
\includegraphics[width=1.1\linewidth]{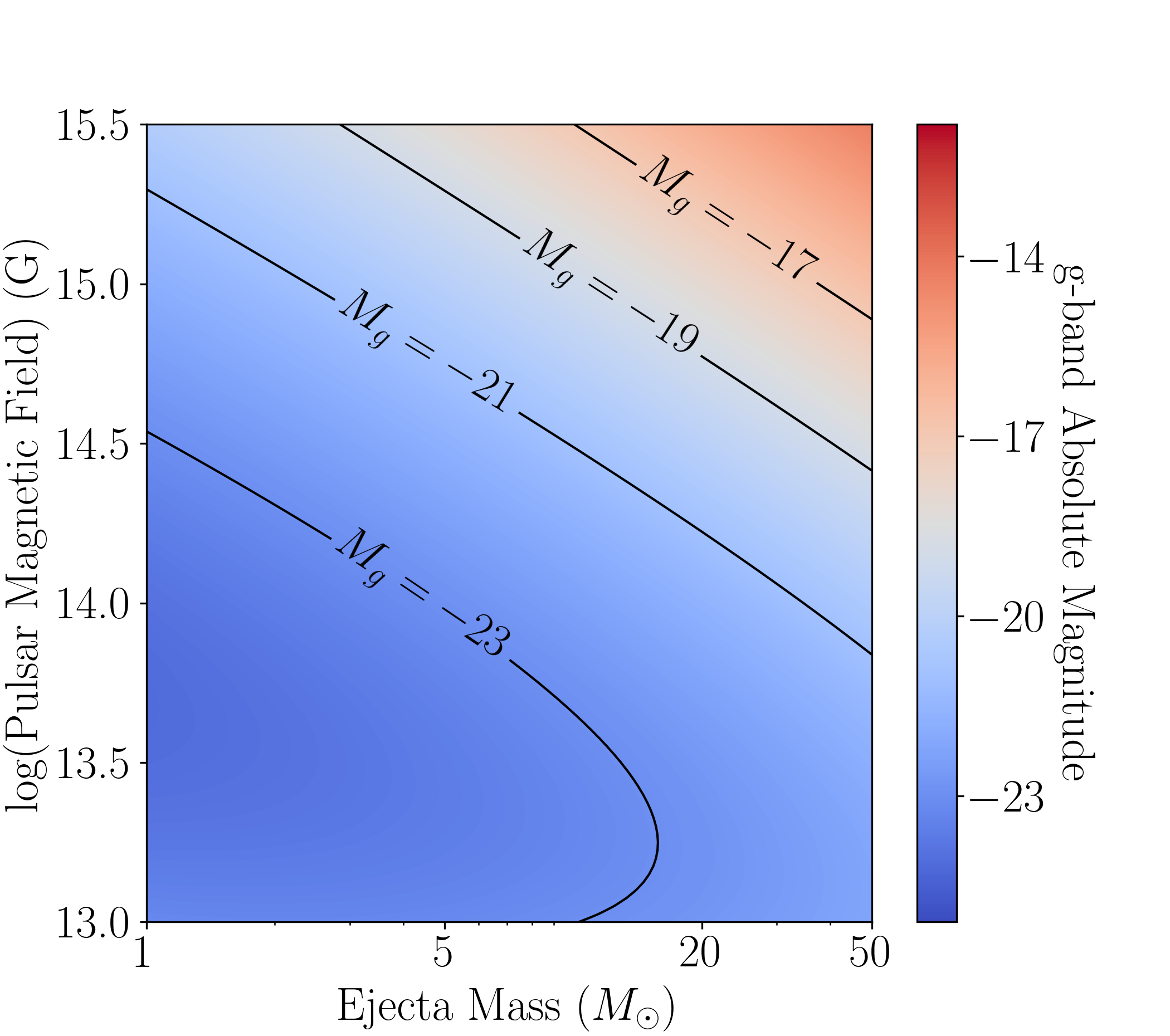}&
\includegraphics[width=1.1\linewidth]{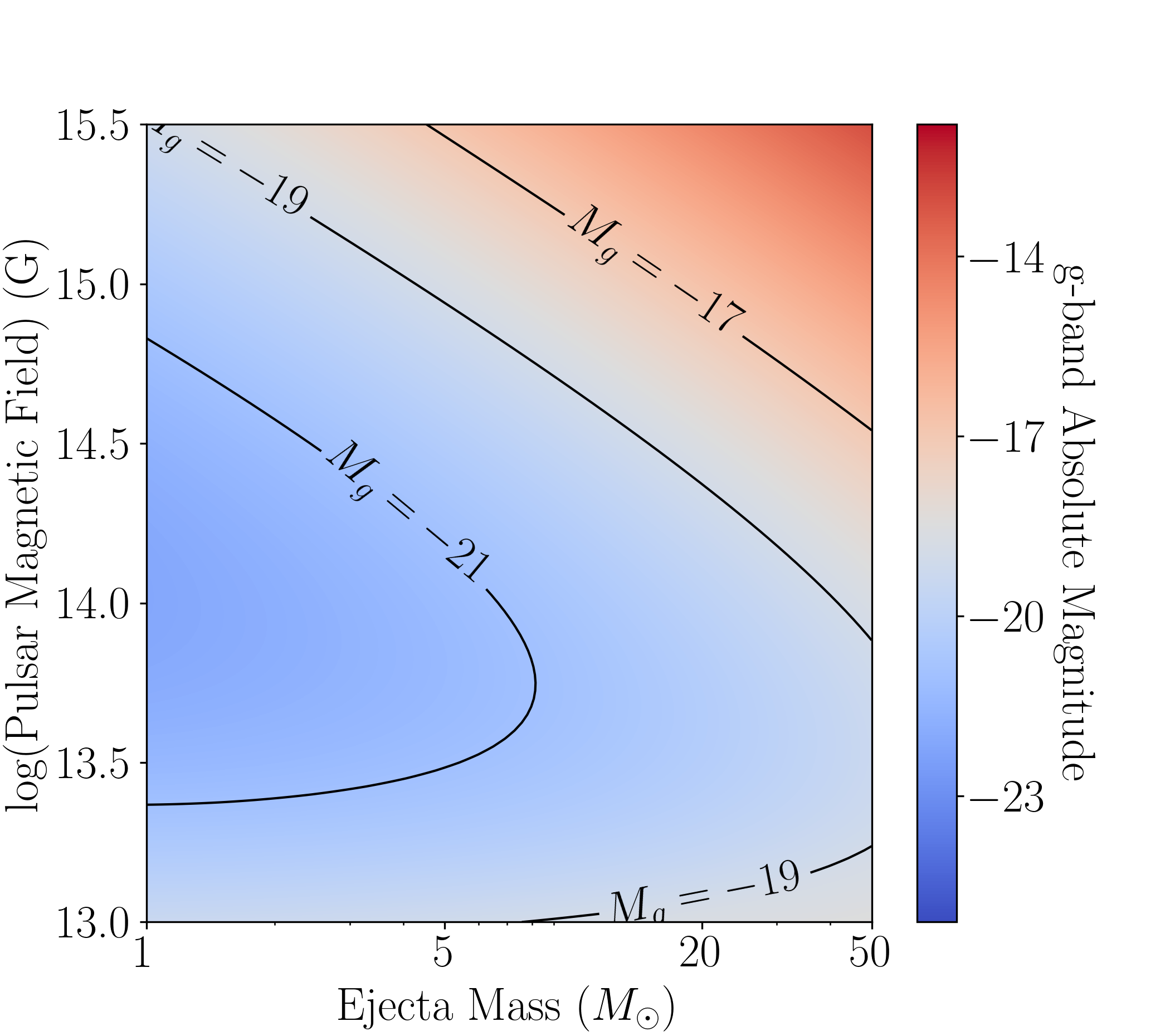}&
\includegraphics[width=1.1\linewidth]{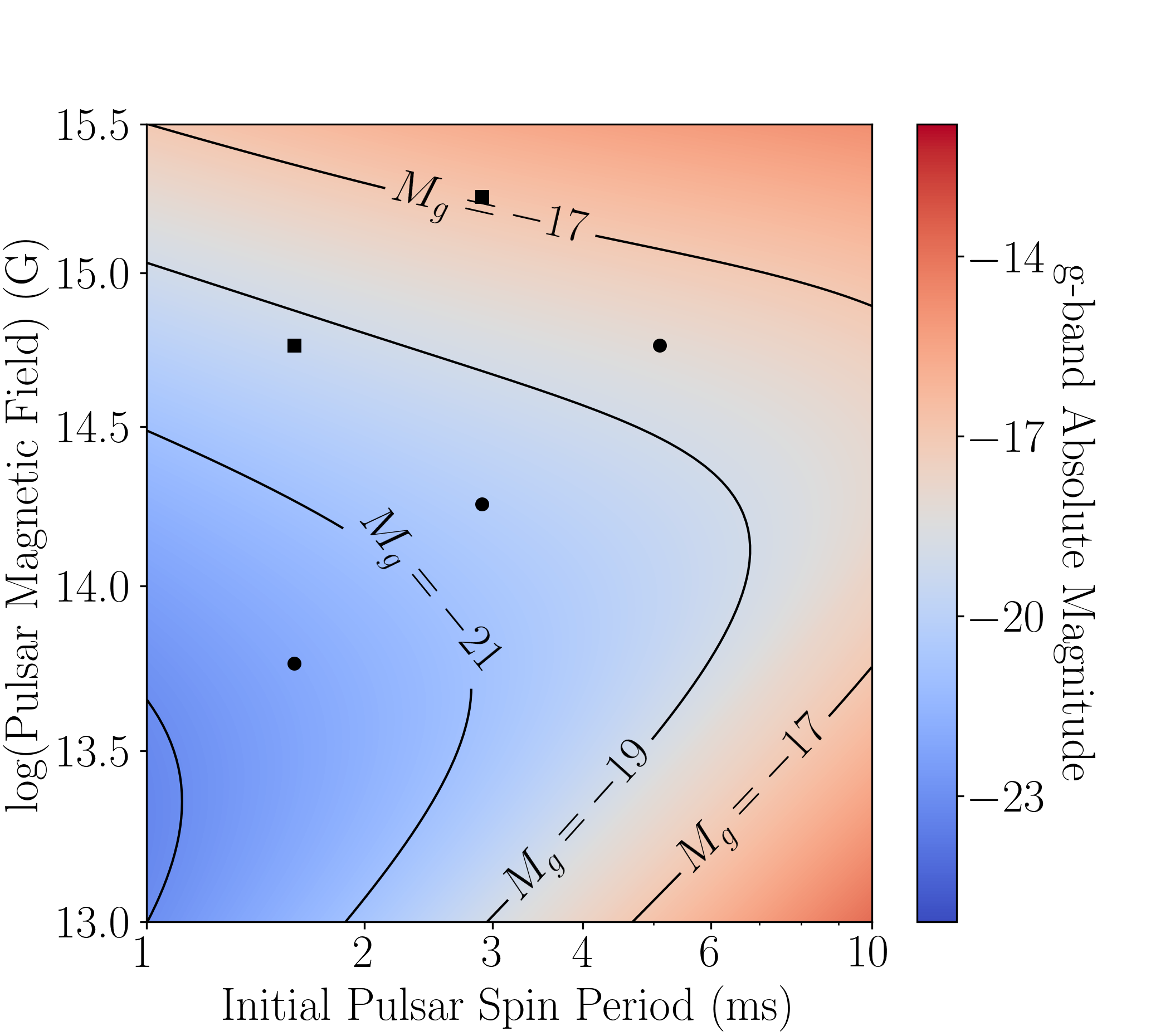}\\[-1.0ex]
\boldsymbol{$g - r$}&
\includegraphics[width=1.1\linewidth]{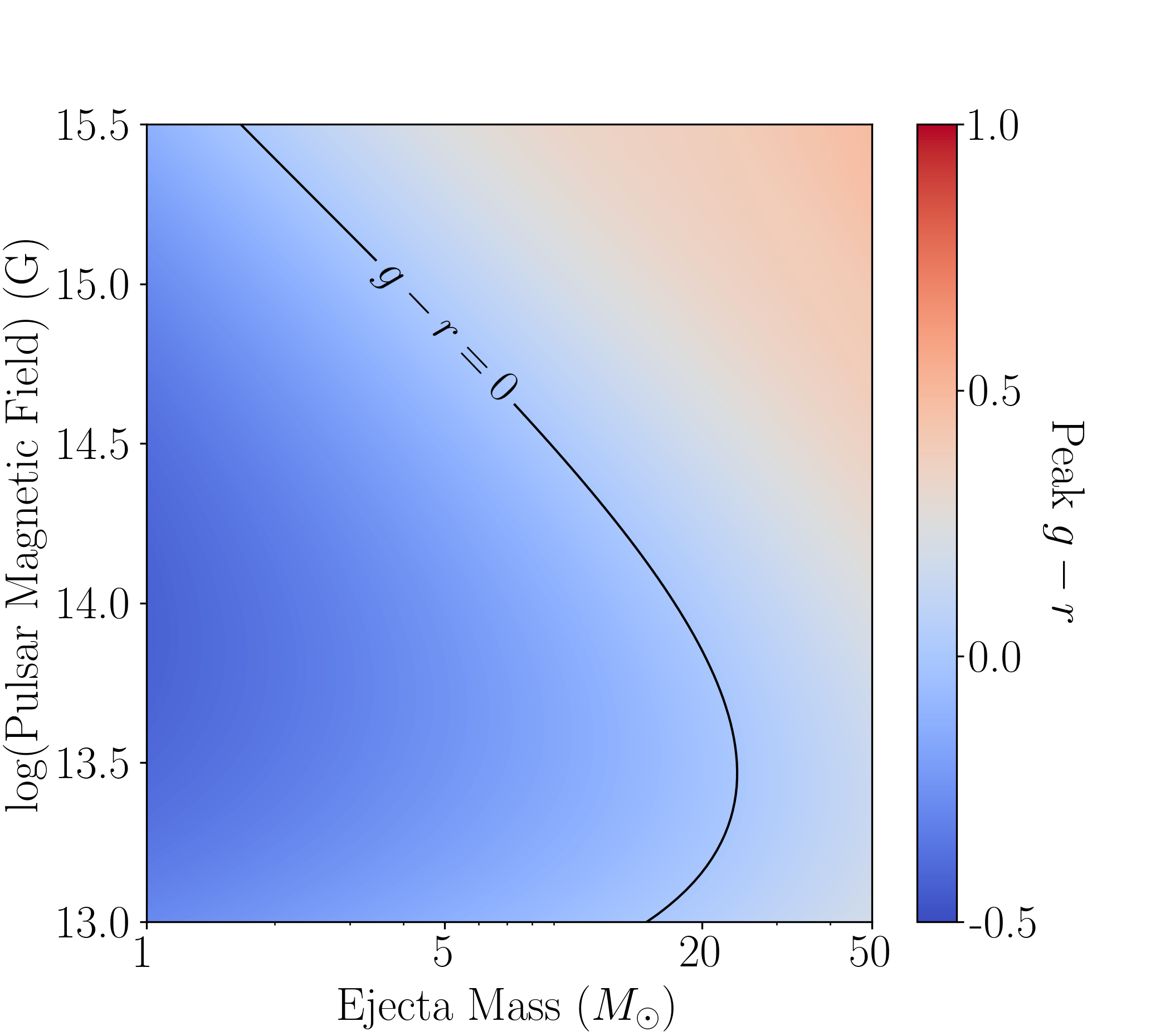}&
\includegraphics[width=1.1\linewidth]{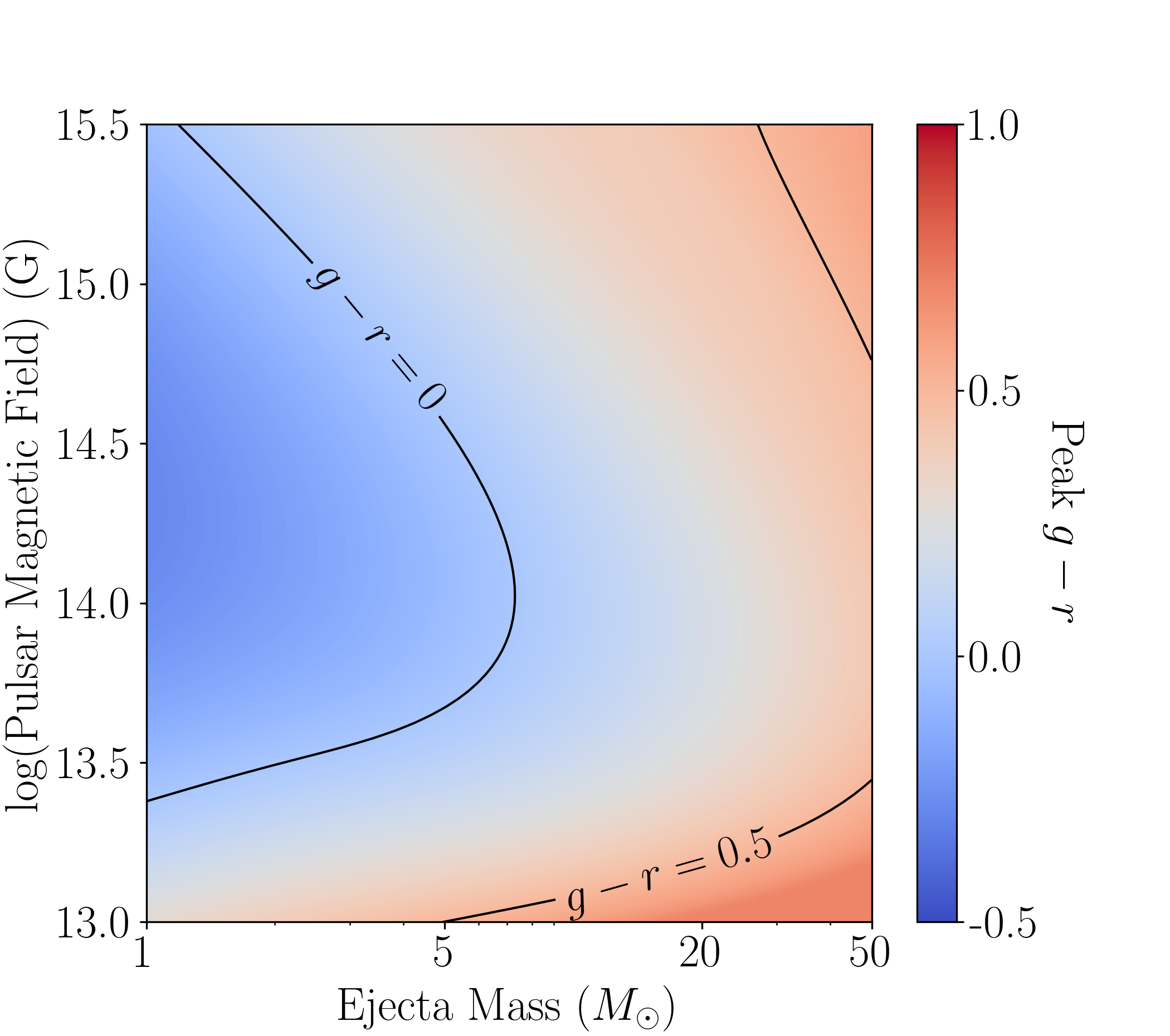}&
\includegraphics[width=1.1\linewidth]{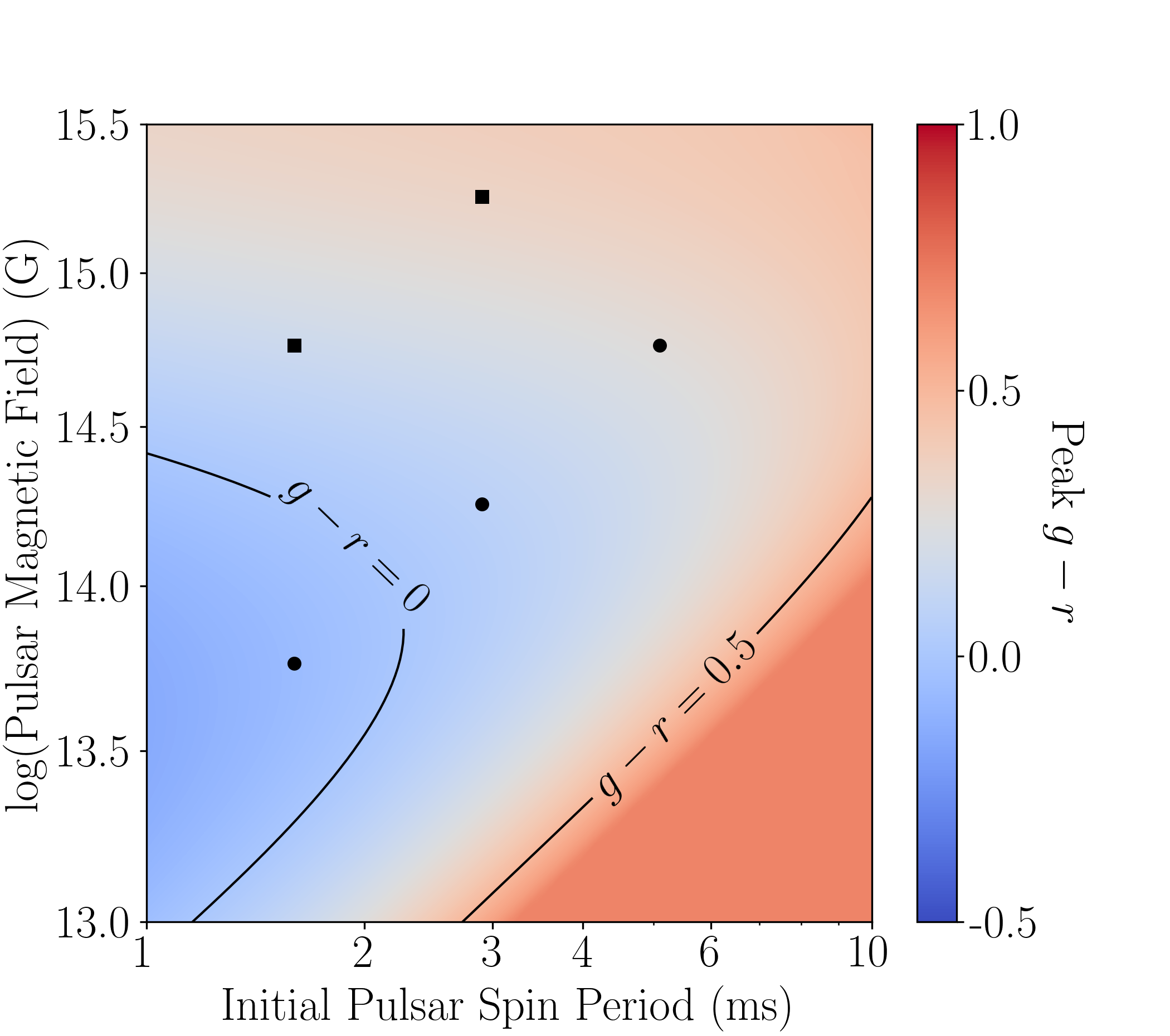}\\[-1.5ex]
\end{tabular}}
\caption{$g$-band absolute magnitude $M_g$ (top) and $g - r$ colour (bottom) at peak for supernovae with varying ejecta mass and $P_0 = 1$ ms (left) and $P_0 = 3$ ms (middle) and with varying spin period and $M_{\rm ej} = 10 M_\odot$ (right).  The black lines indicate notable values of $M_g$ (-17, -19, -21, and -23) and $g - r$ (0 and 0.5). The black symbols indicate the parameters used in \citet{Suzuki2021}, with the circles representing the L46 models and the squares representing the L48 models.}%
\label{fig:ggmr}
\end{figure*}

Figure \ref{fig:ggmr} shows the peak $g$-band absolute magnitude and peak $g - r$ colour over the same parameter grid.  We find a portion of parameter space for low spin period, ejecta mass, and magnetic field which produces a transient more luminous than any previously observed superluminous supernova. 
However, it is unlikely such a combination (particularly low spin period and magnetic fields) can be conceived as magnetic-field amplification mechanisms such as the magneto-rotational instability or Kelvin-Helmholtz instability likely amplify most typical progenitor fields to larger poloidal fields than seen in this parameter space~\citep[e.g.,][]{Reboul-Salze2021}, meanwhile, the stability of magnetic-field configurations in this part of the parameter space is also questionable~\citep[e.g.,][]{Braithwaite2009} and so magnetars in this parameter space may never materialise. 
We note that as we do not track gravitational-wave losses, the newly born magnetar could potentially spin-down rapidly through gravitational-wave radiation~\citep{Sarin2021_review} in this parameter space, depleting the energy reservoir to power such a luminous transient. To compound this all further, it is unknown whether stellar explosions with such small ejecta masses could harbour magnetars that are rapidly rotating but have quite weak poloidal fields in the first place.
The parameter space between $M_g = -21$ and $M_g = -23$, where SLSN are, shifts to lower masses for higher spin periods, where the parameter space around $M_g = -19$, where most SNe Ic-BL are, require either a large ejecta mass ($\gtrsim 5 M_\odot$), higher spin period, or extremely high magnetic field.  The parameter space where $g - r < 0$ mostly overlaps with the $M_g < -21$ region, showing that most SLSNe should have $-0.5 < g - r < 0$ at peak, which is broadly consistent with observations \citep{Chen2023a}.  \citet{Suzuki2021} only present bolometric results, but find an increase in peak bolometric magnitude of $\sim$ 5 in 2D simulations and $\sim$ 4 in 1D simulations between the most and least energetic L46 models, while we find an increase of $\sim$ 4 in $g$-band magnitude over the same parameter difference.  The magnitudes are also roughly consistent with the analytical model presented in \citet{Kashiyama2016}.  The energy ratios from Figure \ref{fig:erat} roughly coincide with the peak $M_g$ for the parameter region where $\zeta \lesssim 1$, but as the spin-down time increases, a larger fraction of the emission is emitted post-peak.  The $g - r$ colours show similar behaviour to $M_g$, but show less dependence on magnetic field.  We speculate that this is because the peak timescale is shorter, and thus the ejecta has had less time to cool.

%\centering%

\subsection{Effect of Varying Braking Index} \label{sec:brake}

\begin{figure*}
\includegraphics[width=0.33\linewidth]{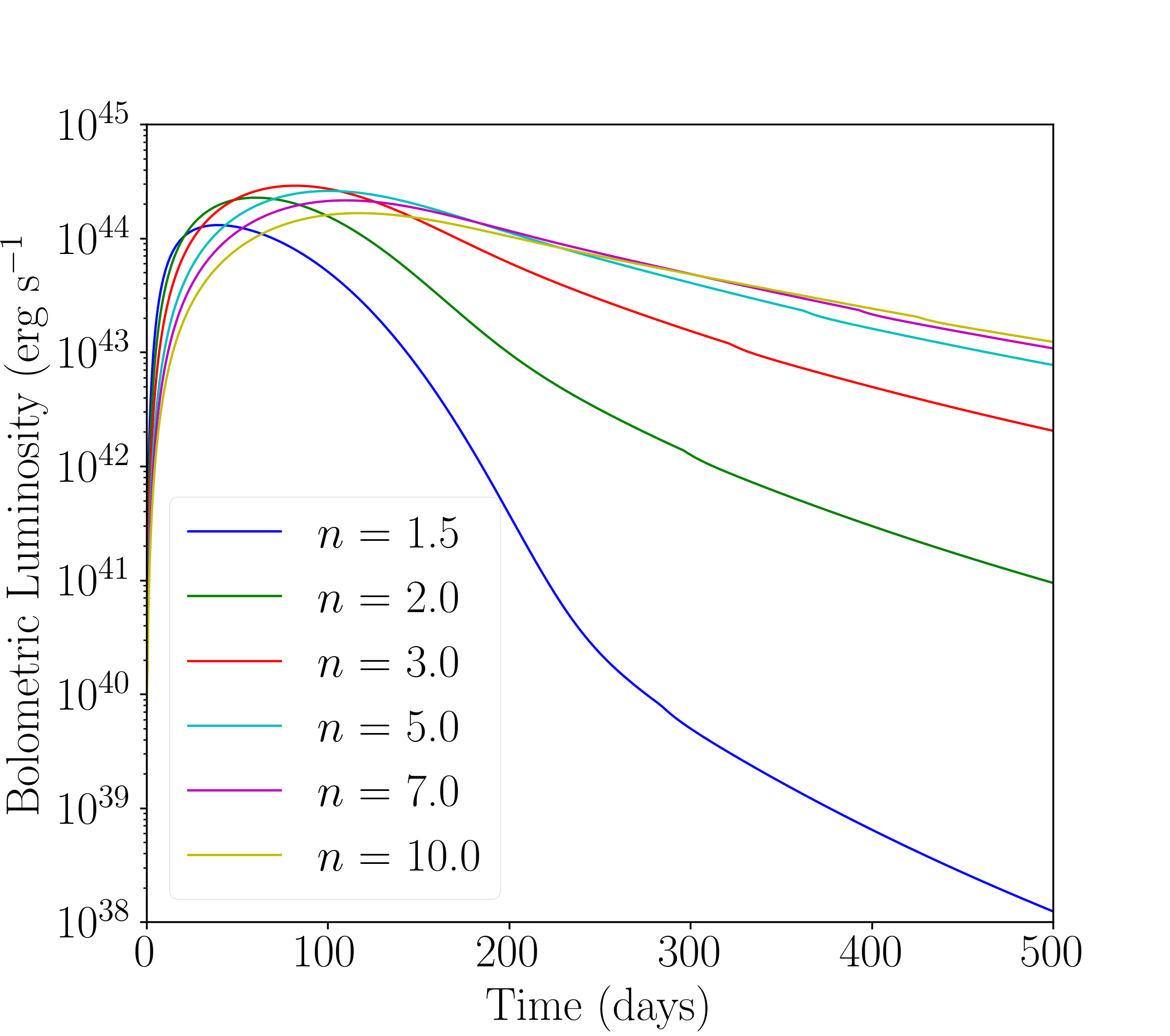}
\includegraphics[width=0.33\linewidth]{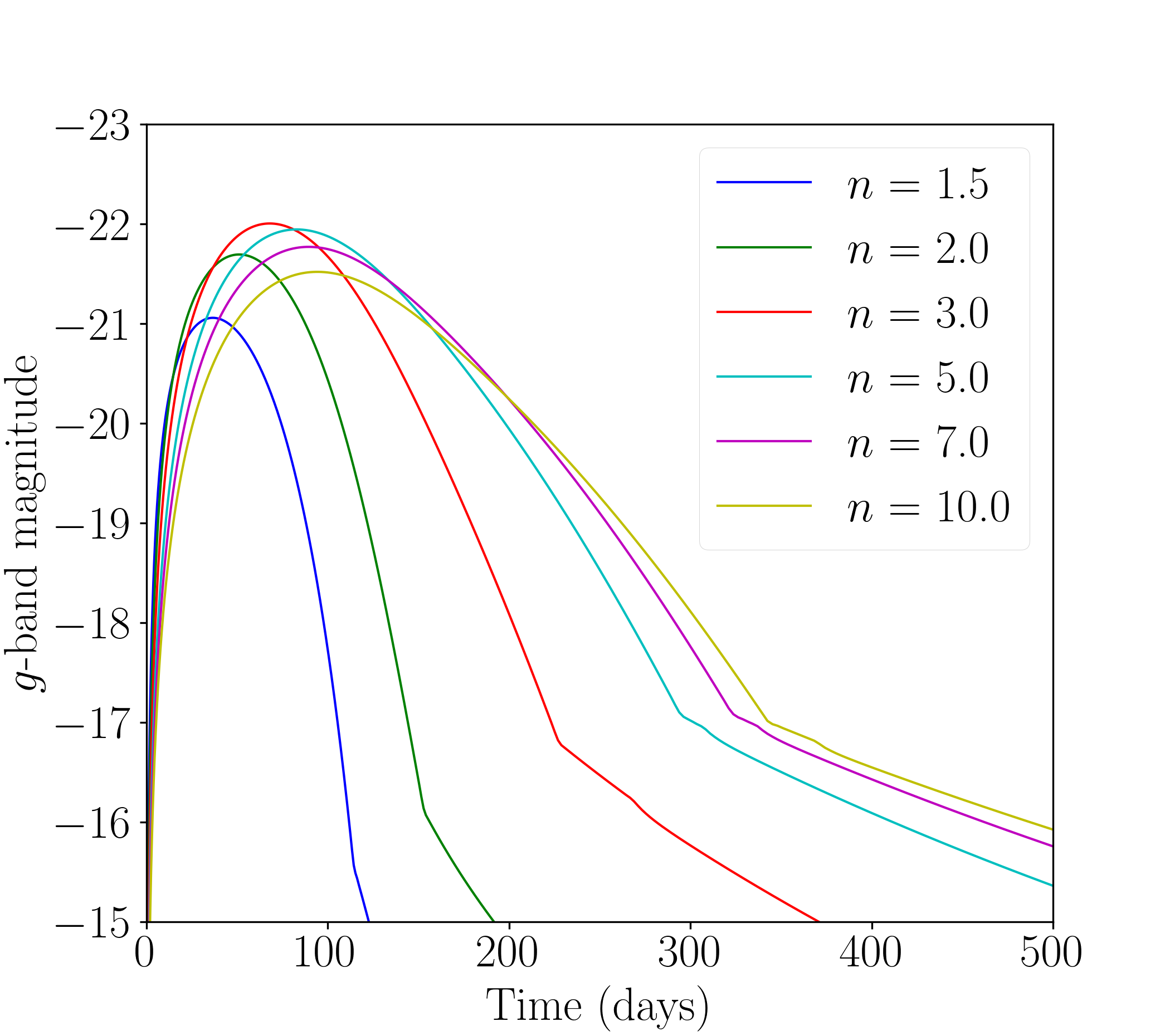}
\includegraphics[width=0.33\linewidth]{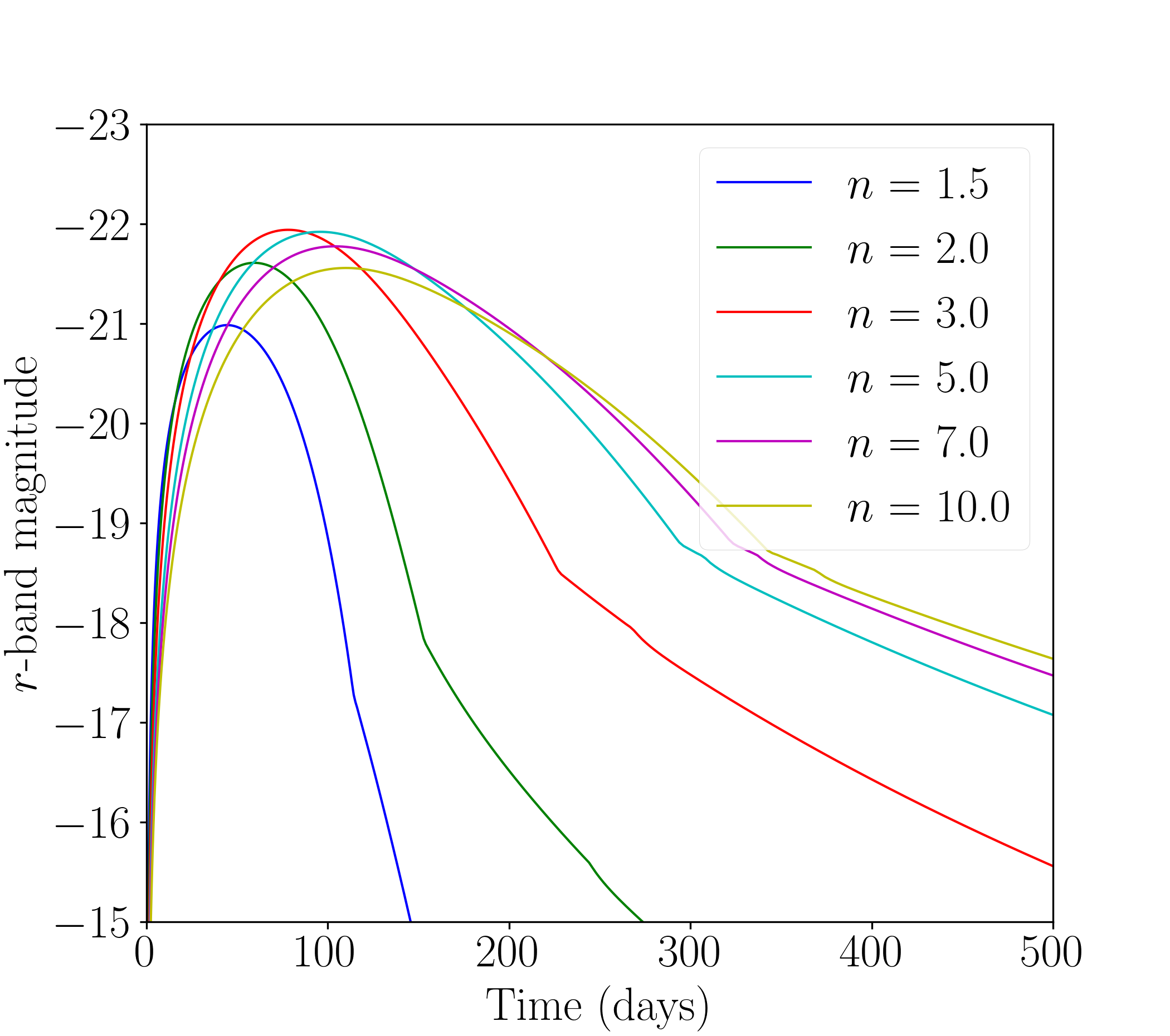}
\caption{Bolometric luminosity (left), absolute $g$-band magnitude (middle), and absolute $r$-band magnitude (right) for several supernovae where only the braking index $n$ is varied.}%
\label{fig:nvar}
\end{figure*}

Light curve luminosity and morphology can vary significantly with variations in magnetar braking index.  Figure \ref{fig:nvar} shows the bolometric luminosity, absolute $g$-band magnitude, and absolute $r$-band magnitude for several supernovae where only the braking index $n$ is varied.  The ejecta mass, spin-down time, and total rotational energy are fixed to 10 $M_\odot$, 10$^6$ s, and 10$^{52}$ erg, with initial magnetar luminosity calculated from Equation \ref{eqn:elt}; all other parameters are the same as in Section \ref{sec:etos}.  The timing of the light curve peak can vary by a factor of $\sim$ 3 and late-time luminosities by orders of magnitude, with large $n$ peaking later and having higher luminosities at later times, although the variation in late-time luminosity asymptotes as $n$ increases due to the exponent in Equation \ref{eqn:llasky} asymptoting to $-1$.  The peak luminosity can vary by $\sim$ 1 mag and is highest for $n \approx$ 3 in this case, although this is not true in general, and these numbers will vary depending on the energetics and diffusion time of the supernova.  The requirement of constant rotational energy drives down $L_0$ (Equation \ref{eqn:elt}), leading to less energy injected at early times and more at later times, which causes the peak luminosity to decreases at large $n$.

\section{Case Studies for Inference} \label{sec:cases}

As a proof of concept, we perform inference on several different classes of supernovae to see if the model is flexible enough to recover sensible parameters for a variety of objects.  First, we validate the model using a simulated SLSN.  Then, we perform inference on SN 2015bn, an SLSN; SN 2007ru, a SN Ic-BL; ZTF20acigmel (better known as ``the Camel''), an FBOT; and iPTF14gqr, a USSN.

Inference is performed on multiband photometry using the open-source software package {\sc{Redback}}~\citep{sarin23_redback} with the {\sc{dynesty}} sampler \citep{Speagle2020} implemented in {\sc{Bilby}} \citep{Ashton2019, Romero-Shaw2020}. We sample with a Gaussian likelihood and an additional white noise term, and sample in flux density rather than magnitude.  
We use the default priors in all cases (shown in Table~\ref{tbl:modparam}) except for the explosion energy of iPTF14gqr, where the lower limit is reduced to 5 $\times$ 10$^{48}$ erg to capture the lower expected explosion energies of USSNe \citep{Suwa2015}, and the plateau temperature of ZTF20acigmel, where the observed photospheric temperature stayed at $\sim$ 20 000 K for the entire time it was detectable in optical/UV \citep{Perley2021}.  We also sample the unknown explosion time with a uniform prior of up to 100 days before the first observation and an extinction term $A_V$ with a uniform prior between 0 and 2, and use a constraint that the total rotational energy of the magnetar $E_{\rm rot} \lesssim 10^{53}$ erg.

\begin{figure*}
\includegraphics[width=0.48\linewidth]{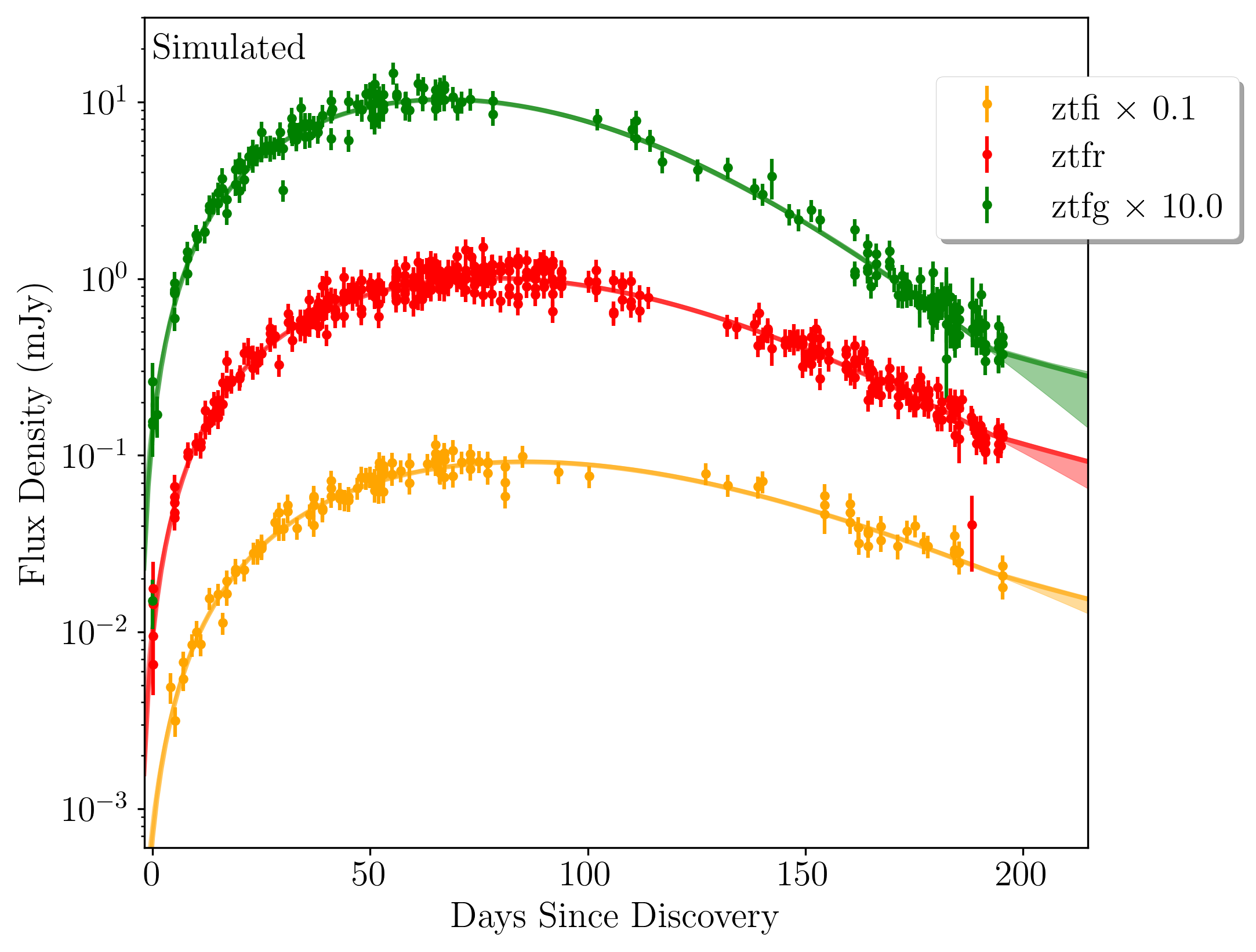}
\includegraphics[width=0.48\linewidth]{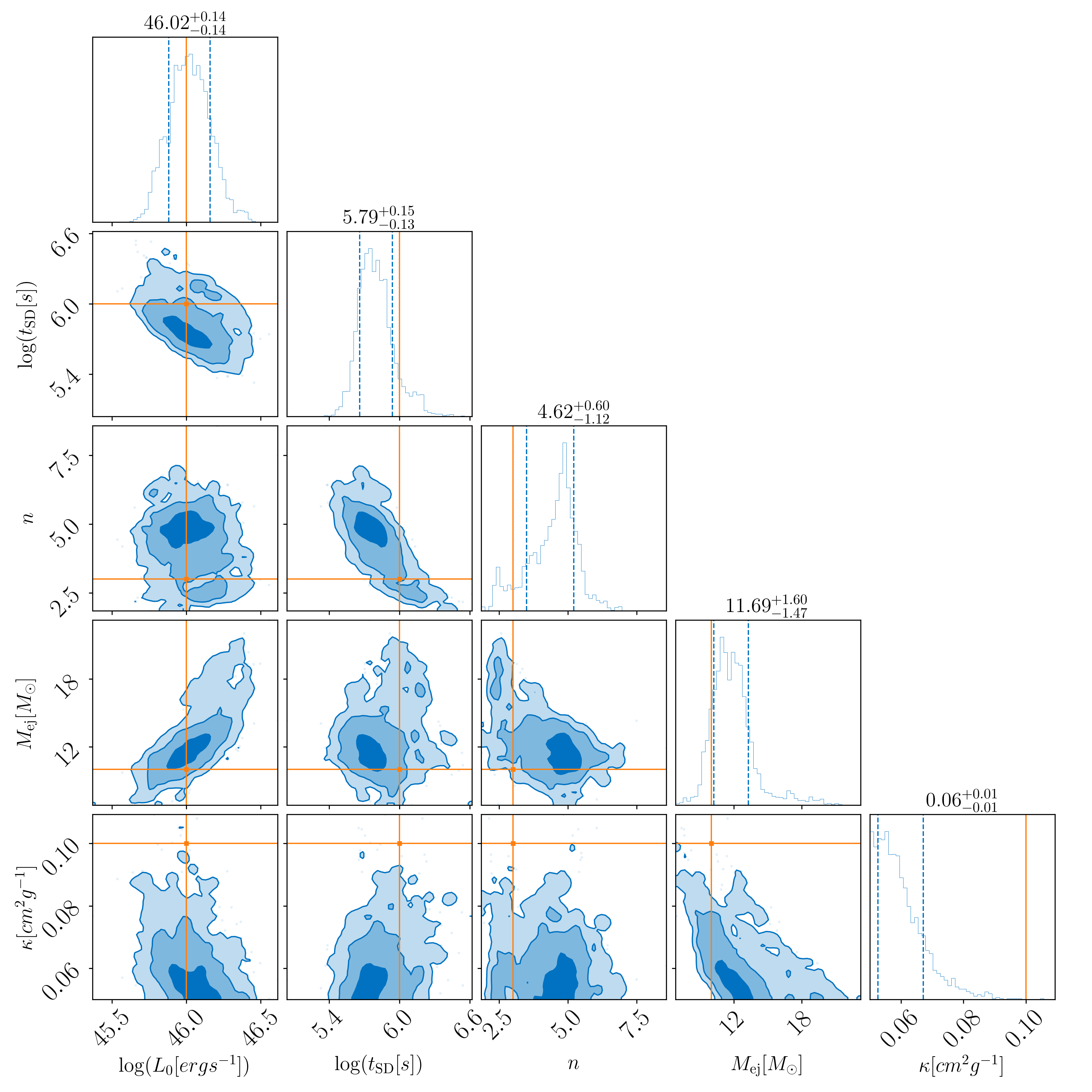}
\caption{Fitted light curve (left) and posteriors of key parameters (right) for the simulated SLSN.  The solid lines in the light curve plot indicate the light curve from the model with the highest likelihood, while the shaded area indicates the 90$\%$ credible interval. The orange dots and lines in the posterior indicate the injected parameters.}%
\label{fig:simver}
\end{figure*}

\begin{figure*}
\includegraphics[width=0.48\linewidth]{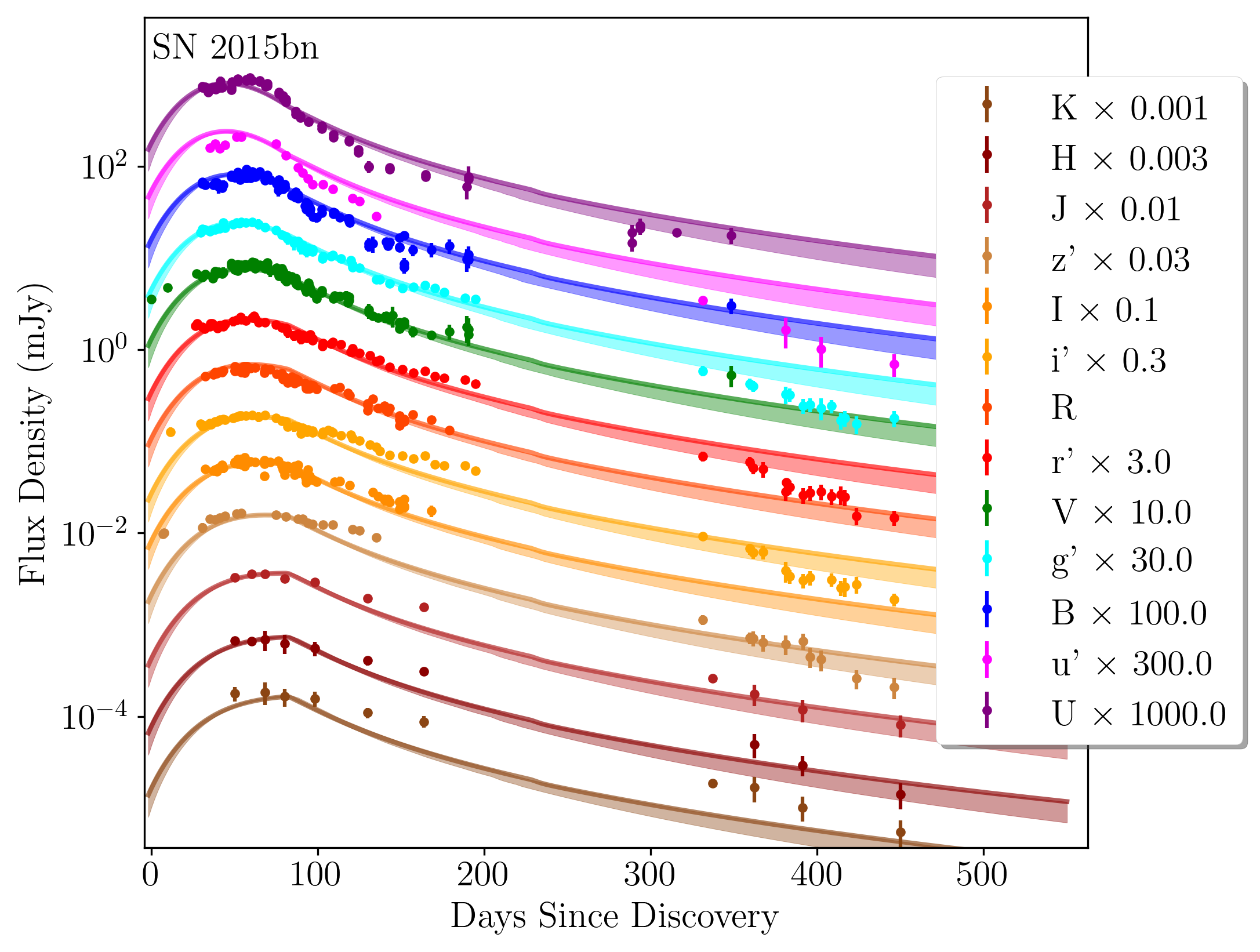}
\includegraphics[width=0.48\linewidth]{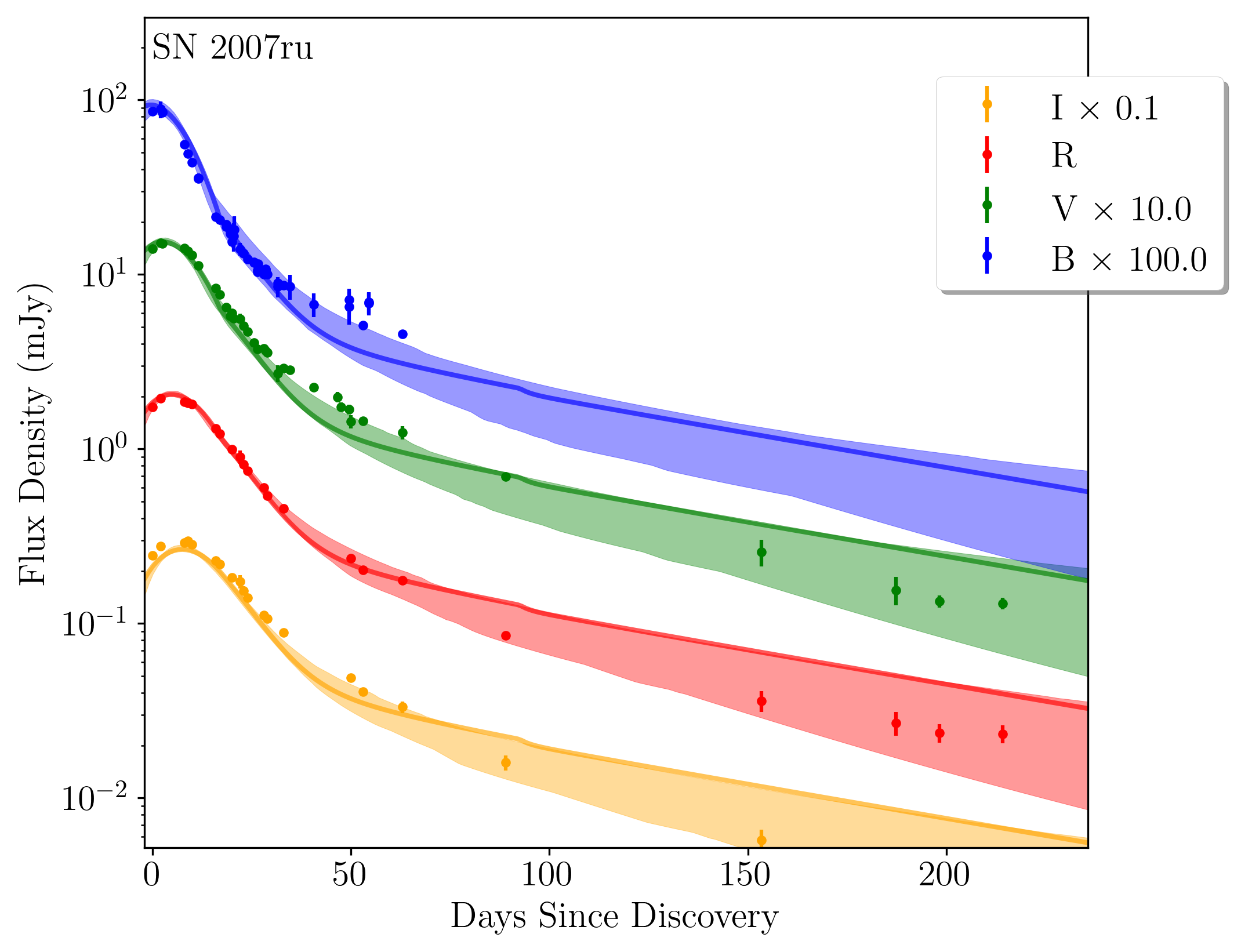} \\
\includegraphics[width=0.48\linewidth]{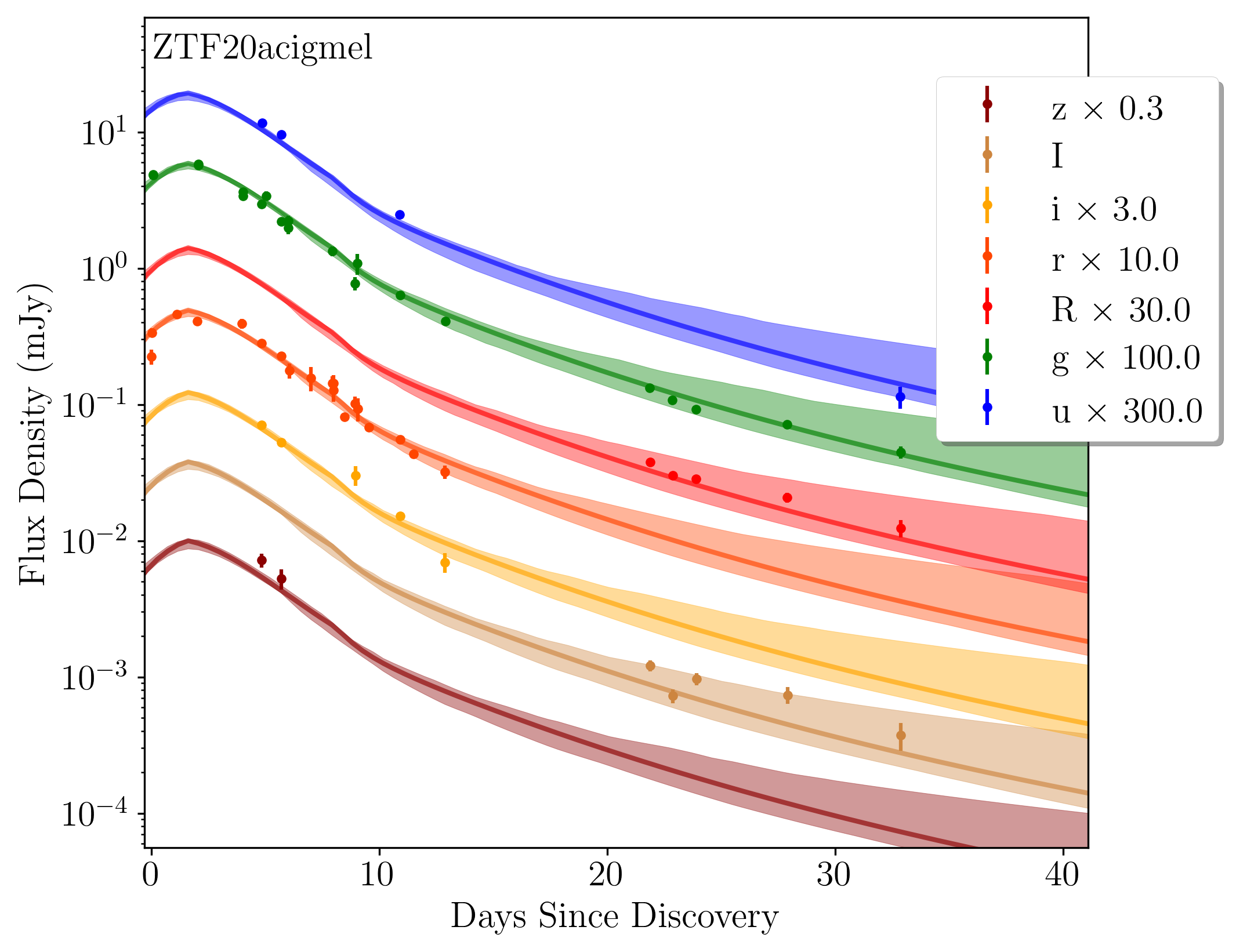}
\includegraphics[width=0.48\linewidth]{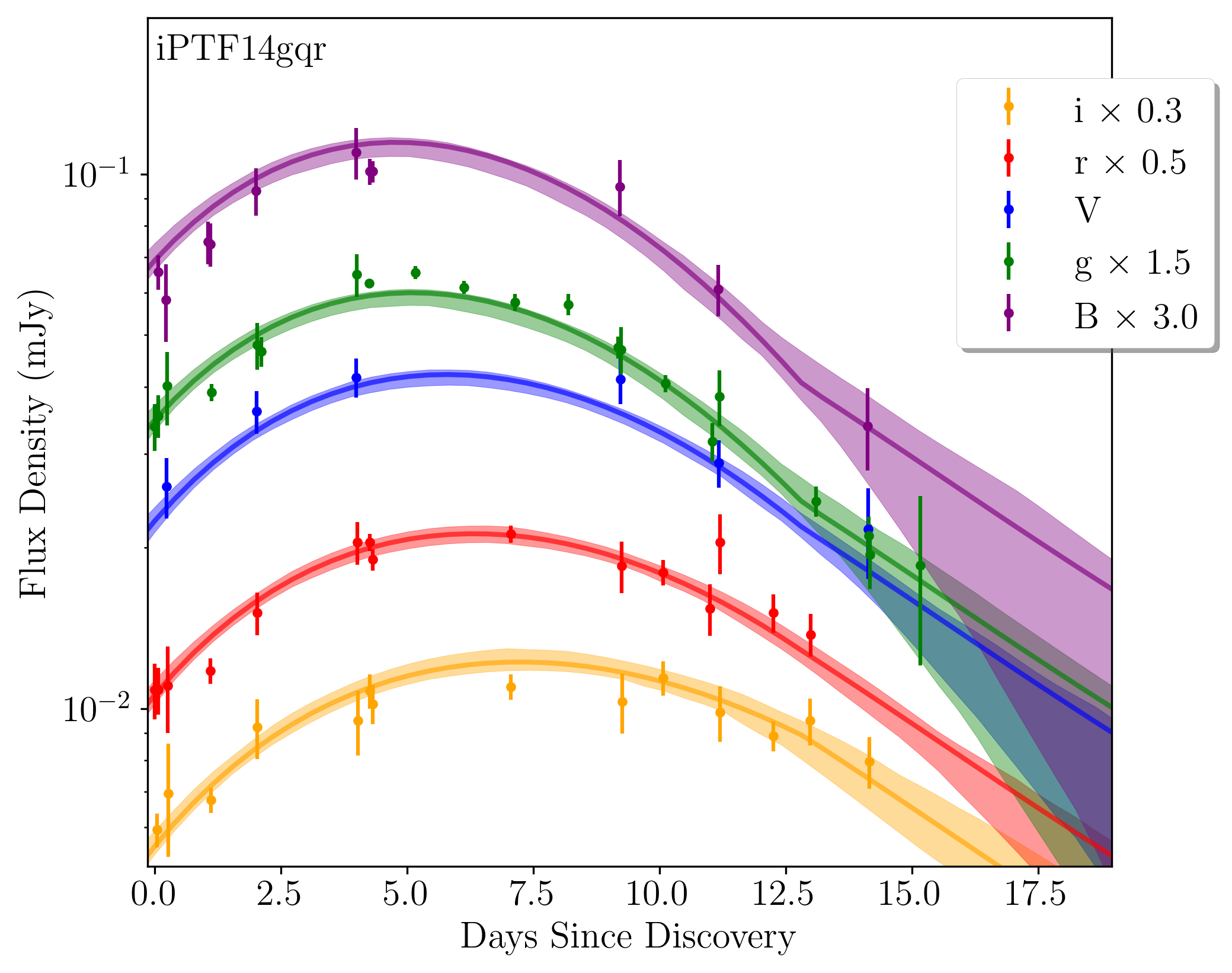}
\caption{Fitted light curves for the four supernovae from our case studies. The solid lines indicate the light curve from the model with the highest likelihood, while the shaded area indicates the 90$\%$ credible interval.}%
\label{fig:lcfits}
\end{figure*}

The fitted light curve and corner plot for the simulated SN are shown in Figure \ref{fig:simver}, the fitted light curves for the other SNe are shown in Figure \ref{fig:lcfits}, the input parameters for the simulated SN and recovered parameters for each SN are shown in Table \ref{tbl:inferredparams}, and the corner plots for each SNe are shown in Appendix \ref{app:corner}.

\begin{table*}
\centering
\begin{tabular}{ccccccc}
   Object & Supernova type & log($L_0$) [erg s$^{-1}$] & log($t_{\rm SD}$) [s] & $n$ & $M_{\rm ej}$ [$M_\odot$] & $\kappa$ [cm$^2$ g$^{-1}$] \\ \hline
   Simulated (Injected) & & 46.0 & 6.0 & 3.0 & 10.0 & 0.1 \\
   Simulated (Recovered) & & 46.02$^{+0.14}_{-0.14}$ & 5.79$^{+0.15}_{-0.13}$ & 4.62$^{+0.60}_{-1.12}$ & 11.69$^{+1.60}_{-1.47}$ & 0.06$^{+0.01}_{-0.01}$ \\ \hline
   SN2015bn & SLSN & 44.81$^{+0.04}_{-0.03}$ & 7.54$^{+0.20}_{-0.21}$ & 6.28$^{+2.43}_{-2.19}$ & 6.05$^{+0.59}_{-0.58}$ & 0.05$^{+0.00}_{-0.00}$ \\
   SN2007ru & Ic-BL & 46.27$^{+0.97}_{-1.02}$ & 3.31$^{+0.78}_{-0.75}$ & 5.24$^{+1.18}_{-1.09}$ & 1.48$^{+0.35}_{-0.33}$ & 0.06$^{+0.02}_{-0.01}$ \\
   ZTF20acigmel & FBOT & 46.15$^{+0.66}_{-0.68}$ & 4.38$^{+0.71}_{-0.49}$ & 7.18$^{+1.53}_{-3.54}$ & 0.20$^{+0.04}_{-0.05}$ & 0.07$^{+0.04}_{-0.02}$ \\
   iPTF14gqr & USSN & 43.31$^{+0.27}_{-0.23}$ & 5.81$^{+0.32}_{-0.35}$ & 4.37$^{+2.14}_{-1.49}$ & 0.15$^{+0.07}_{-0.04}$ & 0.16$^{+0.03}_{-0.02}$ \\   
\end{tabular}
\caption{Injected parameters for the simulated supernovae and median inferred parameter values and 1$\sigma$ uncertainties for the simulated supernova and the four supernovae from the case studies.}
\label{tbl:inferredparams}
\end{table*}

\subsection{Validation on a Simulated Supernova}

To test if the model could recover parameters correctly, we simulated an SLSN with $L_0 = 10^{46}$ erg s$^{-1}$, $t_{\rm SD} = 10^6$ s, $n = 3$, $M_{\rm ej} = 10$ $M_{\odot}$, and $\kappa = 0.1$ cm$^2$ g$^{-1}$ , with the other parameters the same as in Section \ref{sec:result}.  The supernova was placed at redshift $z = 0.1$ and data was generated using the {\sc{Redback}} simulation workflow as observed by ZTF in $g$, $r$ and $i$ band for the first 200 days post explosion.  

The light curve (Figure \ref{fig:simver} (left)) is fit well by the model throughout its evolution.  In the one-dimensional posteriors, only $L_0$ is recovered to within $1\sigma$, while $t_{\rm SD}$, $n$, and $M_{\rm ej}$, are recovered to just outside $1\sigma$ and $\kappa$ is not recovered well at all.  In the two-dimensional posteriors, every parameter is recovered to within $2\sigma$ except for the $n$ - $M_{\rm ej}$ posterior and anything involving $\kappa$.  The correlations between $L_0$, $t_{\rm SD}$, and $n$ also suggests that the rotational energy of the magnetar is well constrained to $\sim$ 10$^{52}$ erg, i.e., the injected value.  The correlation between $M_{\rm ej}$ and $\kappa$ is the main reason they are outside the $1\sigma$ error region in the one-dimensional posteriors.  $n$ also shows an anti-correlation with $\kappa_\gamma$ (see Figure \ref{fig:simulated_corner}), showing that the effect of lowering the braking index can be mimicked in some cases by incomplete gamma-ray thermalization, although this degeneracy can likely be broken by acquiring data at later times.  This result shows the importance of understanding the various opacities of different types of supernovae as well as the necessity of reporting inferred opacity values.

\subsection{SN 2015bn}

SN 2015bn is an SLSN-I at $z = 0.1136$ that was first discovered by the Catalina Sky Survey on 2014 December 23.  It peaked at 79 rest-frame days post discovery, which made it one of the slowest evolving SLSNe at the time, and had peak magnitudes of $M_g = -22.0 \pm 0.08$ mag (AB) and $M_U = -23.07 \pm 0.09$ mag (Vega), making it one of the most luminous as well \citep{Nicholl2016}.  Since then, it has been followed-up extensively in optical/UV/NIR, with photometry and spectroscopy \citep{Nicholl2016, Nicholl2016b, Nicholl2018}, and polarimetry \citep{Inserra2016, Leloudas2017}, and well as in radio \citep{Nicholl2018, Eftekhari2021, Murase2021} and X-rays \citep{Inserra2017, Bhirombhakdi2018}.  SN 2015bn shows strong undulations in the light curve on a timescale of 30-50 days \citep{Nicholl2016, Inserra2017}, and was detectable in optical/UV for more than 1000 days \citep{Nicholl2018}, although the supernova has yet to be detected in either radio or x-rays.

We import the observational data \citep{Nicholl2016, Nicholl2016b} from the Open Supernova Catalog \citep{Guillochon2017}.  Our model is able to fit the supernova peak very well in all bands but underestimates the IR bands in the post-peak photospheric phase and overestimates several bands after 300 days.  The inferred rotational energy of the magnetar is $\sim 6 \times 10^{52}$ erg, which is close to the maximum rotational energy that can be extracted from a newborn magnetar.  The gamma-ray opacity we find is around $10^{-2.4}$ cm$^2$ g$^{-1}$, which is consistent with \citet{Vurm2021} at around 200-300 days, when leakage starts to become noticeable.  The magnetar also shows a large spread in inferred braking indices, but excludes $n = 3$ at $\gtrsim 95\%$ confidence, meaning the vacuum dipole is not a good approximation for this object.  Using Equations \ref{eqn:erot} and \ref{eqn:elt} to infer the spin period, assuming a 1.4 $M_\odot$ neutron star with the same equation of state as \citet{Nicholl2017}, gives $P_0 \approx 0.7$ ms, extremely close to the mass-shedding limit.  %Using the scaling relations from Equations \ref{eqn:l0scale} and \ref{eqn:tsdscale} for a 1.4 $M_\odot$ neutron star with the same equation of state as \citet{Nicholl2017}, we get a spin period $P_0 \approx 0.7$ ms, extremely close to the mass-shedding limit, and a magnetic field $B \approx 8 \times 10^{12}$ G.  

Comparing to the results of \citet{Nicholl2017} shows strong discrepancy between most inferred parameters.  Using Equations \ref{eqn:elt}, \ref{eqn:l0scale}, and \ref{eqn:tsdscale} with parameters from \citet{Nicholl2017} show that they find a spin-down timescale of $\sim$ 105 days, much lower than our $\sim$ 400 days, and a magnetar rotational energy of only $\sim$ 10$^{52}$ erg, a factor of $\sim$ 5 lower than recovered by our model.  This is because the magnetar needs to supply both the radiated and kinetic energy of the supernova in our model, while in the \citet{Nicholl2017} model the ejecta velocity is input separately.  We also recover a smaller ejecta mass of $\sim$ 6 $M_\odot$, compared to their $\sim$ 12 $M_\odot$, and a smaller opacity of $\sim$ 0.05 cm$^2$ g$^{-1}$, compared to their $\sim$ 0.19 cm$^2$ g$^{-1}$.

\subsection{SN 2007ru}

SN 2007ru is an SN Ic-BL at $z = 0.01546$ that was first discovered by the Himalayan Chandra Telescope on 2007 December 2 \citep{Sahu2009}.  The peak magnitude of the supernova was $M_V \approx -19.06$ mag, one of the brighter SNe Ic-BL, and estimated rise time was $8 \pm 3$ days \citep{Sahu2009}.  The photospheric velocity was estimated to be around 20 000 km s$^{-1}$.  A nickel-powered model estimated the $M_{\rm ej} \approx$ 1.3 $M_\odot$ and $M_{\rm Ni} \approx$ 0.4 $M_\odot$ \citep{Sahu2009}, while a previous magnetar-powered model estimated $P_0 \approx 2.30$ ms, $B \approx 6.2 \times 10^{15}$ G, and $M_{\rm ej} \approx 4.43 M_\odot$ \citep{Wang2016}.

We import the observational data \citep{Sahu2009} from the Open Supernova Catalog \citep{Guillochon2017}.  The model fits most of the data well in the optical and NIR. The magnetar energy is $\lesssim 10^{50}$ erg for this supernova, lower than the explosion energy, which dominates the dynamics here.  This energy is also much lower than that estimated by \citet{Wang2016}, and the spin-down time is a factor of $\sim$ 10 larger.  The ejecta mass we estimate is similar to the previous nickel-powered model, but lower than the previous magnetar-powered model.  The opacity we recover is $\sim$ 0.06 cm$^2$ g$^{-1}$, lower than the fixed value of 0.1 cm$^2$ g$^{-1}$ in \citet{Wang2016}.  Finally, the braking index $n$ is much higher than 3, and we can again reject vacuum dipole spin-down at $\gtrsim 95\%$ confidence. Our posterior on the braking index is also consistent with the magnetar spin-down being dominated by gravitational-wave radiation~\citep{Sarin2018, Sarin2020b}. However we caution against strong conclusions based on the measurement of $n$ due to our simplified treatment of the neutron star spin evolution.

\subsection{ZTF20acigmel}

AT2020xnd or ZTF20acigmel, the `Camel', is an FBOT at z = 0.2433 that was first discovered by ZTF \citep{ztf_paper} on 2020 October 12 \citep{Perley2021}.  The peak magnitude was $M_{\rm 5000\AA} \approx -20.6$ mag or $M_{\rm 3900\AA} \approx -20.9$ mag, and estimated rise time was $\sim 2$ days \citep{Perley2021}.  The photospheric radius is already receding at 7 days, and the photospheric temperature from 7-13 days is estimated to be 20 000 $\pm$ 2000 K.  ZTF20acigmel was also found to be luminous in both radio \citep{Ho2022} and x-rays \citep{Bright2022}.

We used the publicly available photometric data from \citet{Perley2021} for our fit.  The model fits the data in all filters throughout the evolution of the object.  The total rotational energy of the magnetar is around 10$^{51}$ erg, comparable to the explosion energy.  Using Equations \ref{eqn:erot} and \ref{eqn:elt} to infer the spin period, again assuming a 1.4 $M_\odot$ neutron star with the same equation of state as \citet{Nicholl2017}, gives $P_0 \approx 5$ ms. This value, as well as the ejecta mass, are consistent with values found for the FBOT distribution \citep{Liu2022}.  The braking index has a wide distribution with two peaks, and is anti-correlated with the gamma-ray opacity, like with the simulated supernova.  If the spin-down mechanism is vacuum dipole, then the gamma-ray opacity is likely $\gtrsim$ 1 cm$^2$ g$^{-1}$, which can happen at late times for PWNe with magnetization similar to Galactic PWNe \citep{Tanaka2010, Tanaka2013, Vurm2021}.  If the braking index is larger, then the gamma-ray opacity is smaller, and the magnetization must be lower, similar to SN 2015bn and SN 2017egm \citep{Vurm2021}.  This can potentially tested by observing non-thermal emission in the late phase.

%The braking index is consistent with vacuum dipole, and using the scaling relations and the same 1.4 $M_\odot$ neutron star as for SN 2015bn, we can derive an initial spin period of $\sim$ 5 ms and magnetic field of $\sim 2 \times 10^{15}$ G; these values, as well as the ejecta mass, are consistent with values found for the FBOT distribution \citep{Liu2022}.

\subsection{iPTF14gqr}

iPTF14gqr is a USSN at $z = 0.063$ that was first discovered by the intermediate Palomar Transient Factory (iPTF) \citep{Law2009} on 2014 October 14. The supernova featured a bright first peak that faded within a day, followed by a more extended light curve the rises after about $\sim$ 4 days.  The first peak can be explained by shock cooling \citep{De2018} and the second peak by either radioactivity \citep{De2018} or a new born magnetar \citep{Sawada2022}, although both peaks can also be explained by interaction alone \citep{Khatami2023}.

We used the publicly available photometric data from \citet{De2018}, although we exclude all data points within one day post-explosion, since we only claim that the second peak could be magnetar powered.  The model fits the data in all filters throughout the evolution of the object.  The explosion energy, which has a wider prior to account for the low explosion energy of USSNe, is constrained to $\sim 6 \times 10^{49}$ erg, while the total magnetar rotational energy is $\sim 2 \times 10^{49}$ erg.  The braking index is not very well constrained; although it is consistent with vacuum dipole to within error.  The median spin-down time of $\approx$ 7 days is slightly lower than that found by \citet{Sawada2022}, although our magnetar energy is a factor of $\sim$ 5 higher; this is likely because our explosion energy is smaller. The ejecta mass we derive is also consistent with estimates by both \citet{De2018} and \citet{Sawada2022}.

\section{Discussion and Summary} \label{sec:conc}

As shown by both the exploration of the parameter space (Section \ref{sec:result}) and the case studies (Section \ref{sec:cases}), the model presented here is incredibly versatile.  Three of the four case studies had previous been fit by a magnetar model \citep{Nicholl2017, Wang2016, Sawada2022}, but each of these studies used a different model with different assumptions to model a particular supernova or class of supernovae.  A versatile model allows comparisons of different populations to be done self-consistently and determine what model variations manifest in vastly different types of supernovae, as well as probe whether a continuum between these sources could exist or whether there are multiple distinct classes.

\begin{figure}
\includegraphics[width=\linewidth]{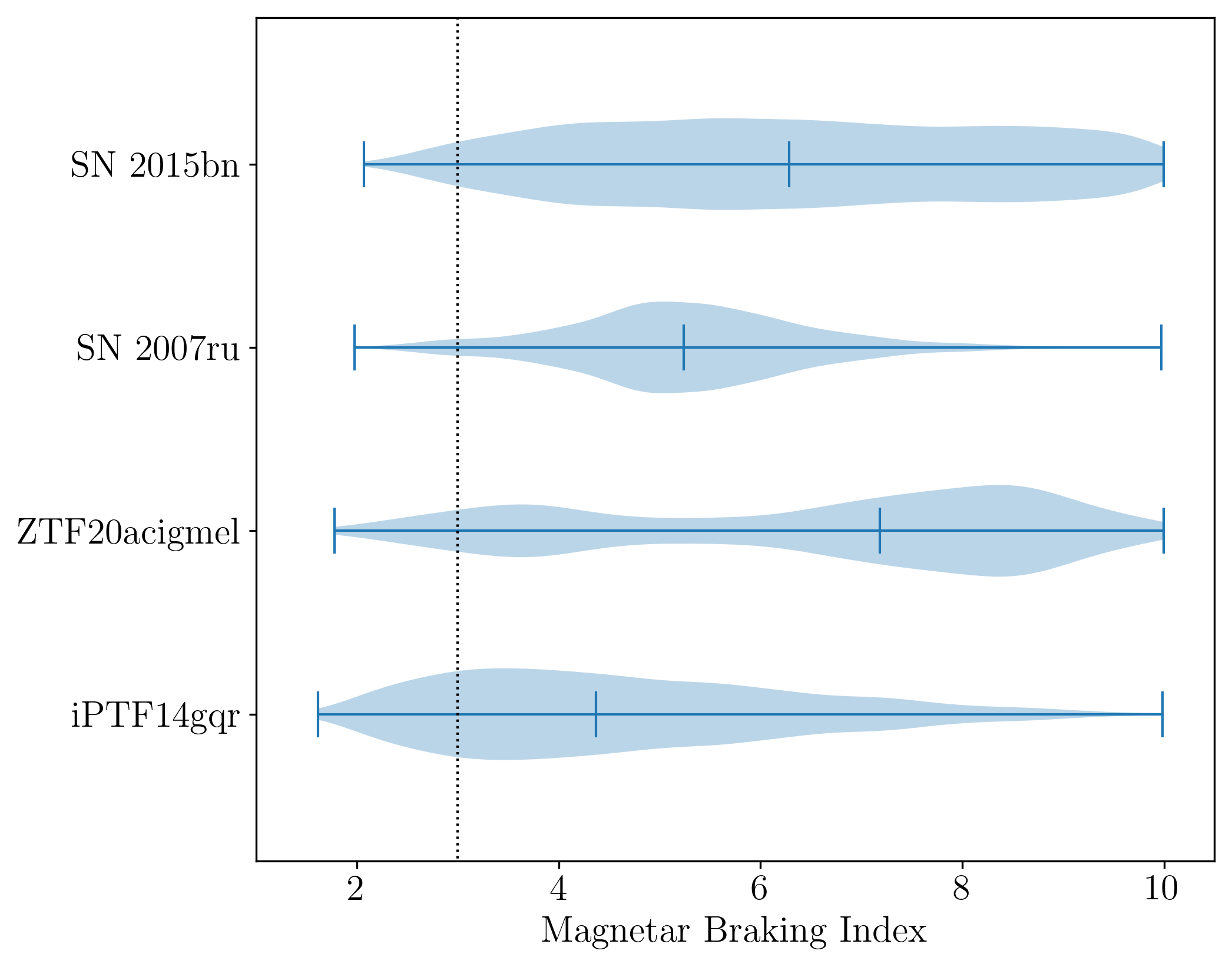} \\
\includegraphics[width=\linewidth]{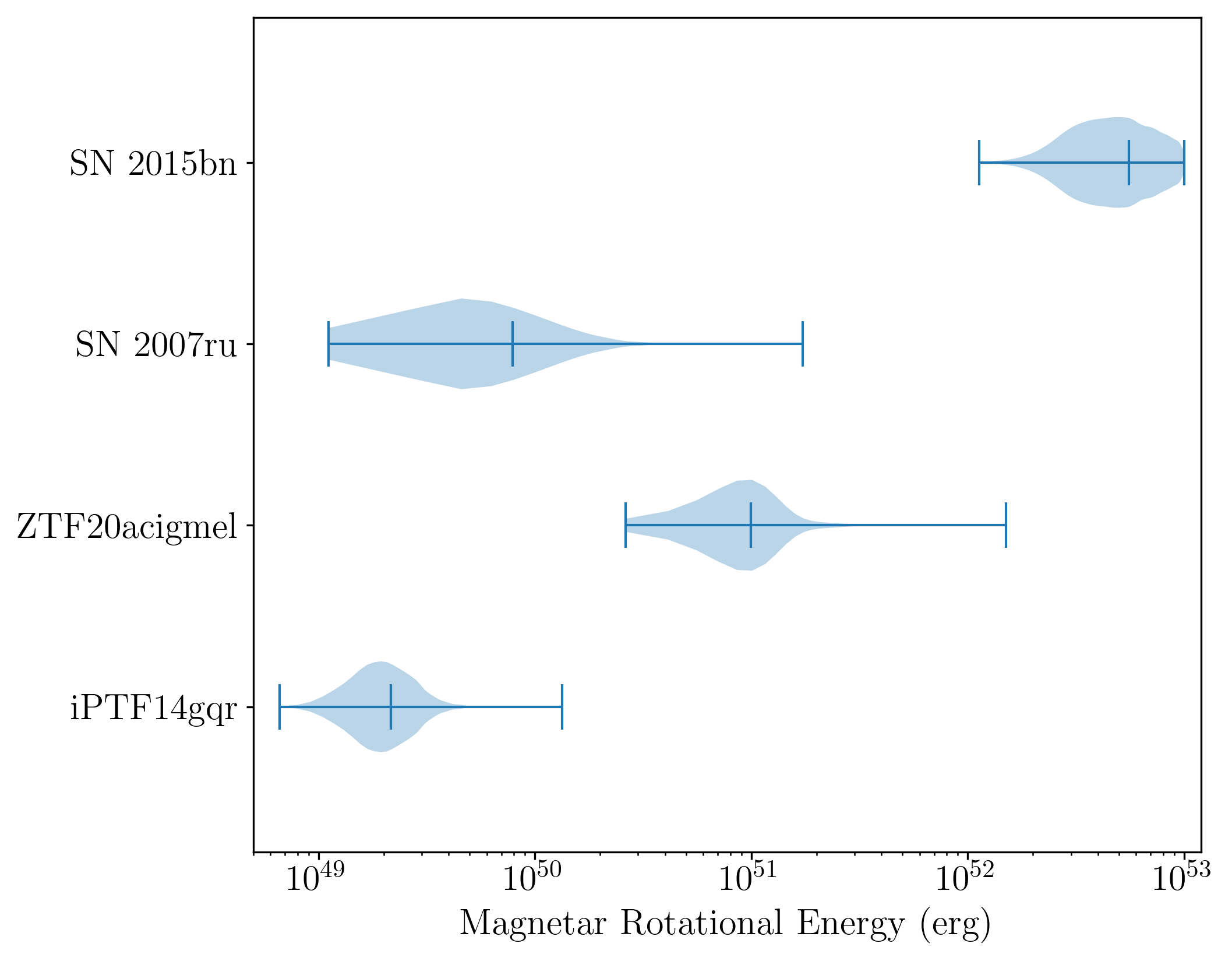} \\
\caption{Posterior distributions of the magnetar braking index (top) and rotational energy (bottom) for the four supernovae from our case studies. The median values are indicated by the blue vertical lines within the distribution.  The black dashed line indicates vacuum dipole spin-down ($n = 3$).}%
\label{fig:violin}
\end{figure}

Much of the flexibility of our model comes from self-consistent dynamical evolution of the ejecta, while the addition of magnetar braking index as a parameter allows for some possible insight into the spin-down mechanism of the newborn millisecond magnetar.  While inferring the spin-down mechanism for any particular supernova is difficult due to the number of possible mechanisms that could be in effect, we can constrain this parameter at a population level and see if different classes have evidence for different spin-down mechanisms.  Figure \ref{fig:violin} (top) shows the posterior probability distribution of braking index for the four case study supernova.  As mentioned above, $n = 3$ is rejected for SN 2015bn and SN 2007ru at $> 95\%$ confidence; SN 2007ru shows a posterior that is roughly Gaussian while SN 2015bn shows a posterior that is roughly uniform at $n \gtrsim 4$.  
ZTF20acigmel and iPTF14gqr both show non-Gaussianity in their posteriors as well; the Camel has a double peaked distribution due to the degeneracy with gamma-ray opacity, while iPTF14gqr has a tail at high $n$ but peaks very close to $n = 3$.  While making definitive statements about spin-down mechanisms will require a much larger sample, this small sample already shows diversity in their inferred braking index, highlighting a potential interesting question for future studies.

All the objects studied in Section \ref{sec:cases}, including the simulated supernova, show a strong negative correlation between $L_0$ and $t_{\rm SD}$.  This shows that the magnetar rotational energy $E_{\rm rot}$, and thus the total energy budget of the supernova, can be constrained for these objects to within an order of magnitude (see Figure \ref{fig:violin} (bottom)).  The ejecta masses were all found to be similar to previous studies, except for SN 2015bn, and the spin-down times were found to differ for all models, usually by a factor of $\sim$ 3-10.  The magnitude of these discrepancies seems to vary depending on the supernova, and a large-scale sample study on supernovae that have been previously characterized by a magnetar model \citep[e.g.][]{Chen2023b} is necessary to characterize the systematic difference between our model and previous models.% be very close to previous models for SN 2015bn and iPTF14gqr \citep{Nicholl2017, Wang2016, Sawada2022}, although the total magnetar energy budget was different in each case due to the way we treated our dynamics.  

Each magnetar-driven transient model \citep[e.g][]{Yu2013, Wang2016, Kashiyama2016, Nicholl2017, Sarin2022, Sawada2022} has slight differences in assumptions.  All of these models assume vacuum dipole spin-down, although \citet{Kashiyama2016} and \citet{Sarin2022} also add a spin-down component for gravitational waves from magnetic deformation.  For gamma-ray thermalization, \citet{Yu2013} uses a constant gamma-ray thermalization efficiency, \citet{Kashiyama2016} models the non-thermal emission and deposition using several additional free parameters, while \citet{Wang2016}, \citet{Nicholl2017}, and \citet{Sarin2022} use the same prescription we do.  \citet{Nicholl2017} and \citet{Sawada2022} do not treat the effect of the magnetar on the ejecta dynamics, while \citet{Wang2016} uses a simplified treatment.  The kilonova models \citep{Yu2013, Sarin2022} have different radioactive heating rates due to the r-process material in the ejecta, while the other supernova models \citep{Wang2016, Kashiyama2016, Nicholl2017, Sawada2022} are non-relativistic and can not be used for transients with exceptionally low ejecta masses and powerful magnetar engines.  Finally, only \citet{Nicholl2017} and \citet{Sarin2022} were originally implemented into a publicly-available Bayesian inference code \citep{Guillochon2017, sarin23_redback}, although the \citet{Yu2013} model has also been implemented in {\sc{Redback}}.

This model has a few caveats which may prevent it from properly describing certain transients.  The first is that it is a one-zone, one-dimensional model.  This makes the treatment of the photospheric radius very simplified compared to real supernovae.  Engine-driven supernovae also show hydrodynamic instabilities in multidimensional simulations \citep[e.g.][]{Chen2016, Chen2020, Suzuki2017, Suzuki2019, Suzuki2021} which can shred the inner ejecta, causing a decrease in the effective optical depth of the ejecta, as well as affecting the timescale for non-thermal leakage.  While our model can qualitatively produce the same behaviour as radiation hydrodynamic simulations, multi-dimensional effects cause the supernova peak to be earlier and more luminous.  If the spin-down timescale of the magnetar is smaller than the Kelvin-Helmholtz timescale of the magnetar for neutrino emission $t_{\rm KH,\nu} \lesssim 100$ s, then baryon loading on the magnetized wind via the neutrino-driven wind can be relevant \citep{Thompson2004} and the magnetized wind can be collimated by anisotropic and hoop stress \citep{Bucciantini2007, Bucciantini2008} and form a jet \citep{Kashiyama2016}.  This model also does not self-consistently track gravitational-wave emission and how it depletes the overall rotational energy reservoir. The model also has no way to explain the bumps and undulations that have recently found in a large number of SLSNe \citep{Hosseinzadeh2022, Chen2023b}, which have been explained by both circumstellar material \citep[e.g.][]{West2023, Chugai2023} and magnetars \citep{Chugai2022, Moriya2022, Dong2023}.  The model we present can only explain a smooth light curve with minimal fine structure; however, this structure is likely connected to small mass ejections or binary interactions in the final few years of the life of the progenitor and may not be strongly connected to the power source of the supernova.  Furthermore, The SED used in our model has a simplified treatment of line blanketing that is calibrated to SLSNe \citep{Nicholl2017}, and may not be accurate for other transients. Although it is possible to switch the SED to a blackbody if line blanketing is not expected to be strong, the SED is also not a good approximation deep into the nebular phase on the supernova, since the emission will start to be dominated by line emission instead of photospheric emission \citep[e.g.,][]{Schulze2023}.

Although the magnetar-driven supernova model is versatile enough to fit light curves of many different supernovae, the best way to determine whether a magnetar is really the power source is to compare the nebular spectra, polarization, and non-thermal emission from the supernovae with different models.  Within the magnetar model, this emission is form the PWN or its interaction with the supernova ejecta.  If the ejecta is optically thin, the synchrotron and inverse Compton emission from the PWN can leak through and be detected \citep{Kotera2013, Metzger2014, Murase2015, Omand2018}, while if the ejecta is optically thick, these photons will be absorbed and change the temperature and electronic state of the ejecta, both giving detectable signals \citep{Chevalier1992, Omand2019, Omand2023}.  However, interaction with circumstellar material can also produce non-thermal emission, polarization, and spectra with high ionization lines, therefore detailed modeling is necessary to make any strong conclusions.

In this work, we present a more flexible, inference-capable, publicly available model for the light curves of magnetar-driven supernovae.  This model can be applied to any transient where the dominant power sources are either a magnetar or $^{56}$Ni, as long as that energy release is roughly spherically symmetric.  The main changes from previous models are the coupling of the magnetar energy injection to the kinetic energy of the supernova and the addition of the magnetar braking index as a free parameter, allowing the exploration of non-vacuum-dipole spin down.  We show that the model can reproduce the basic properties on several phenomenologically different supernovae, and also fit four different types of supernovae, retrieving parameters consistent with works using separate models.  This model will allow us to explore the full diversity of these supernovae, better characterize these supernovae from just their light curves, and make better predictions future multiwavelength emission to better test the magnetar-driven scenario.

\section*{Acknowledgements}
We thank the anonymous referee for their helpful comments.  We also thank Claes Fransson and Steve Schulze for helpful comments and discussions.  N. Sarin is supported by a Nordita Fellowship, Nordita is funded in part by NordForsk. 

%%%%%%%%%%%%%%%%%%%%%%%%%%%%%%%%%%%%%%%%%%%%%%%%%%
\section*{Data Availability}

Light curve data for SN2015bn and SN2007ru was obtained from the Open Supernova Catalog \citep{Guillochon2017} using {\sc{Redback}}.  Light curve data for ZTF20acigmel and iPTF14gqr were downloaded from repositories released with the corresponding papers \citep{Perley2021, De2018}. The model is available for public use within {\sc{Redback}}~\citep{sarin23_redback}.



\bibliographystyle{mnras}
\bibliography{ref} % if your bibtex file is called example.bib

%\begin{thebibliography}{99}
%\end{thebibliography}

%%%%%%%%%%%%%%%%%%%%%%%%%%%%%%%%%%%%%%%%%%%%%%%%%%

%%%%%%%%%%%%%%%%% APPENDICES %%%%%%%%%%%%%%%%%%%%%

\appendix

\section{Posteriors from Case Studies} \label{app:corner}

\begin{figure*}
\includegraphics[width=\linewidth]{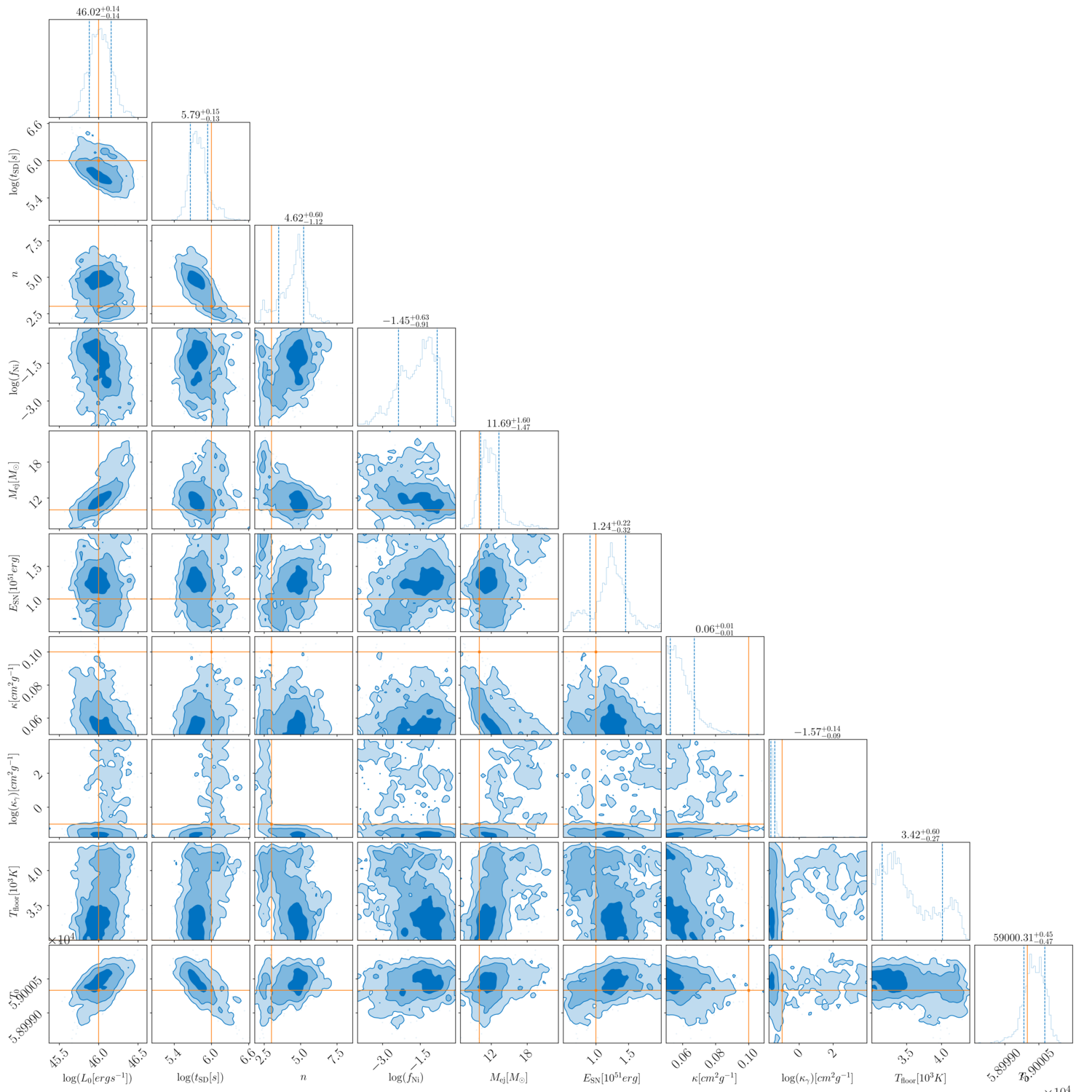}
\caption{Corner plot for the simulated SLSN.  Injected values are shown in orange.}%
\label{fig:simulated_corner}
\end{figure*}

\begin{figure*}
\includegraphics[width=\linewidth]{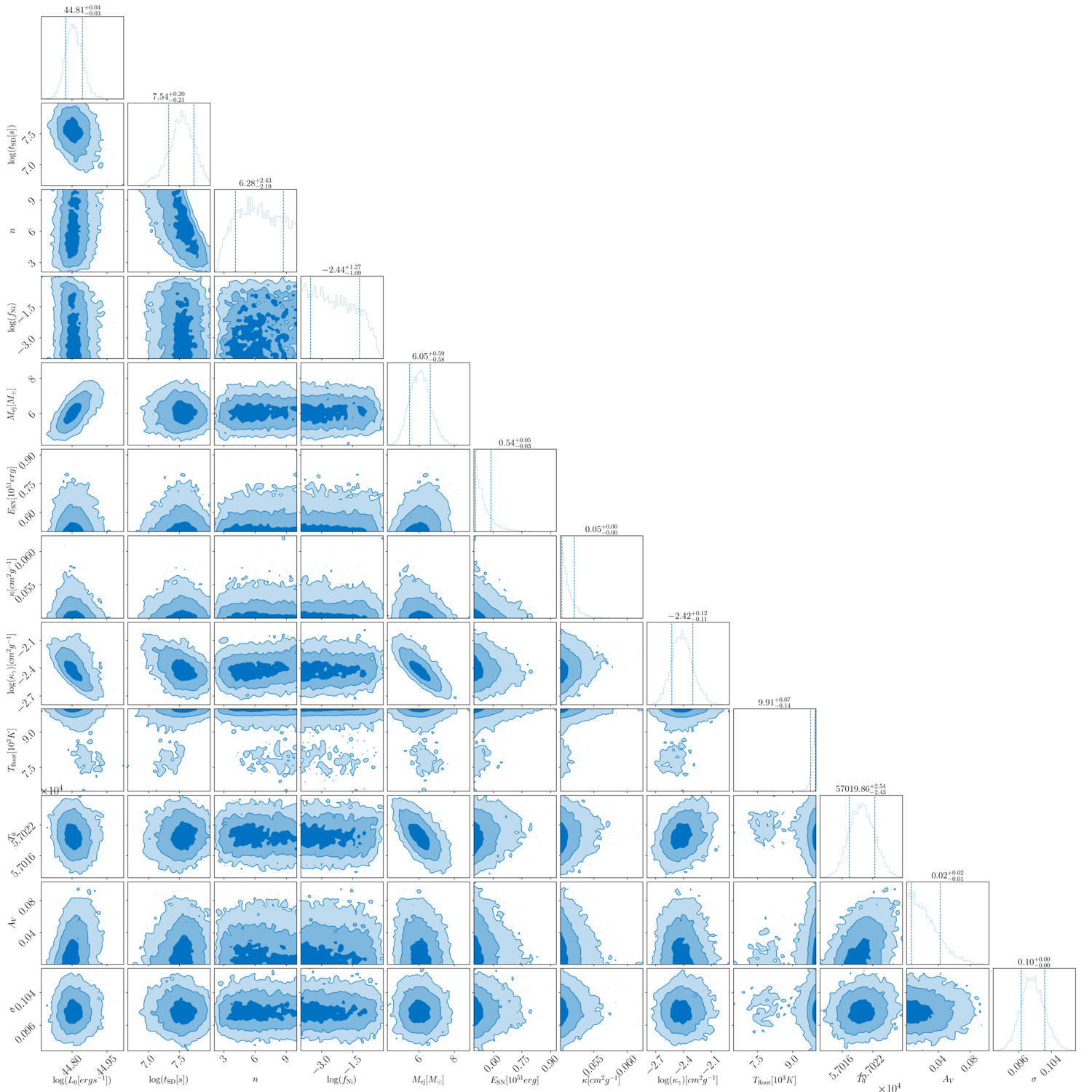}
\caption{Corner plot for SN 2015bn, an SLSN.}%
\label{fig:15bn_corner}
\end{figure*}

\begin{figure*}
\includegraphics[width=\linewidth]{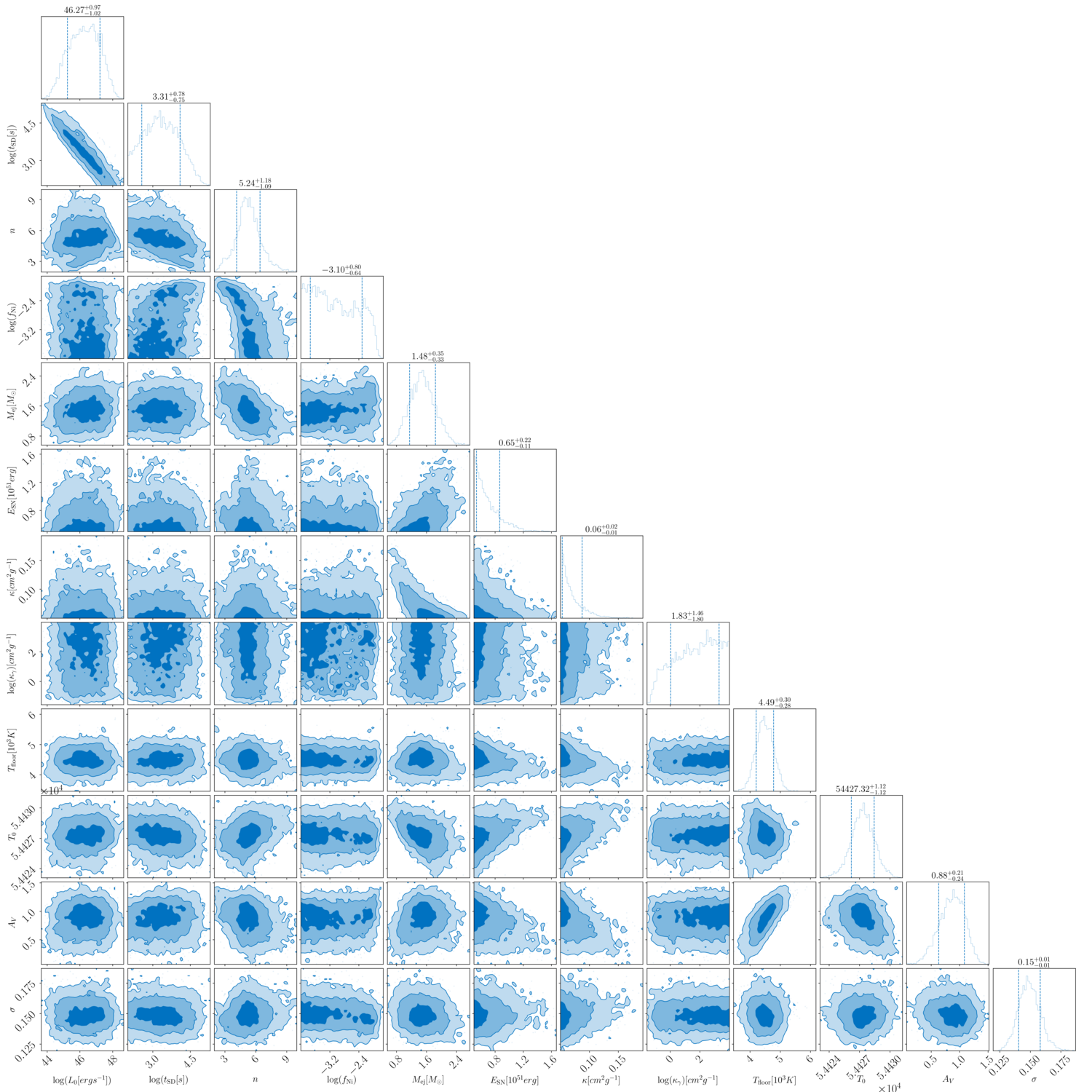}
\caption{Corner plot for SN 2007ru, an SN Ic-BL.}%
\label{fig:07ru_corner}
\end{figure*}

\begin{figure*}
\includegraphics[width=\linewidth]{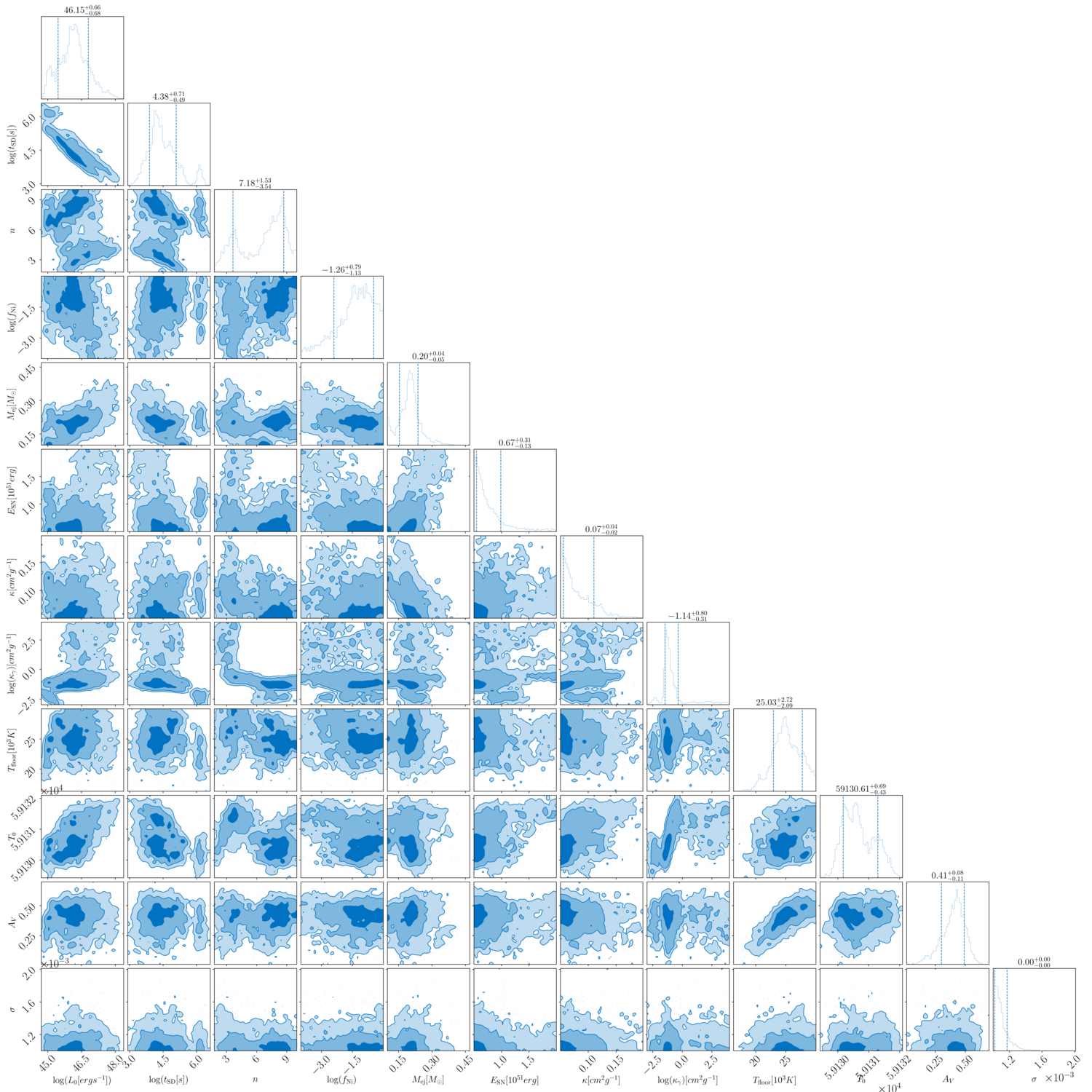}
\caption{Corner plot for ZTF20acigmel (the Camel), an FBOT.}%
\label{fig:camel_corner}
\end{figure*}

\begin{figure*}
\includegraphics[width=\linewidth]{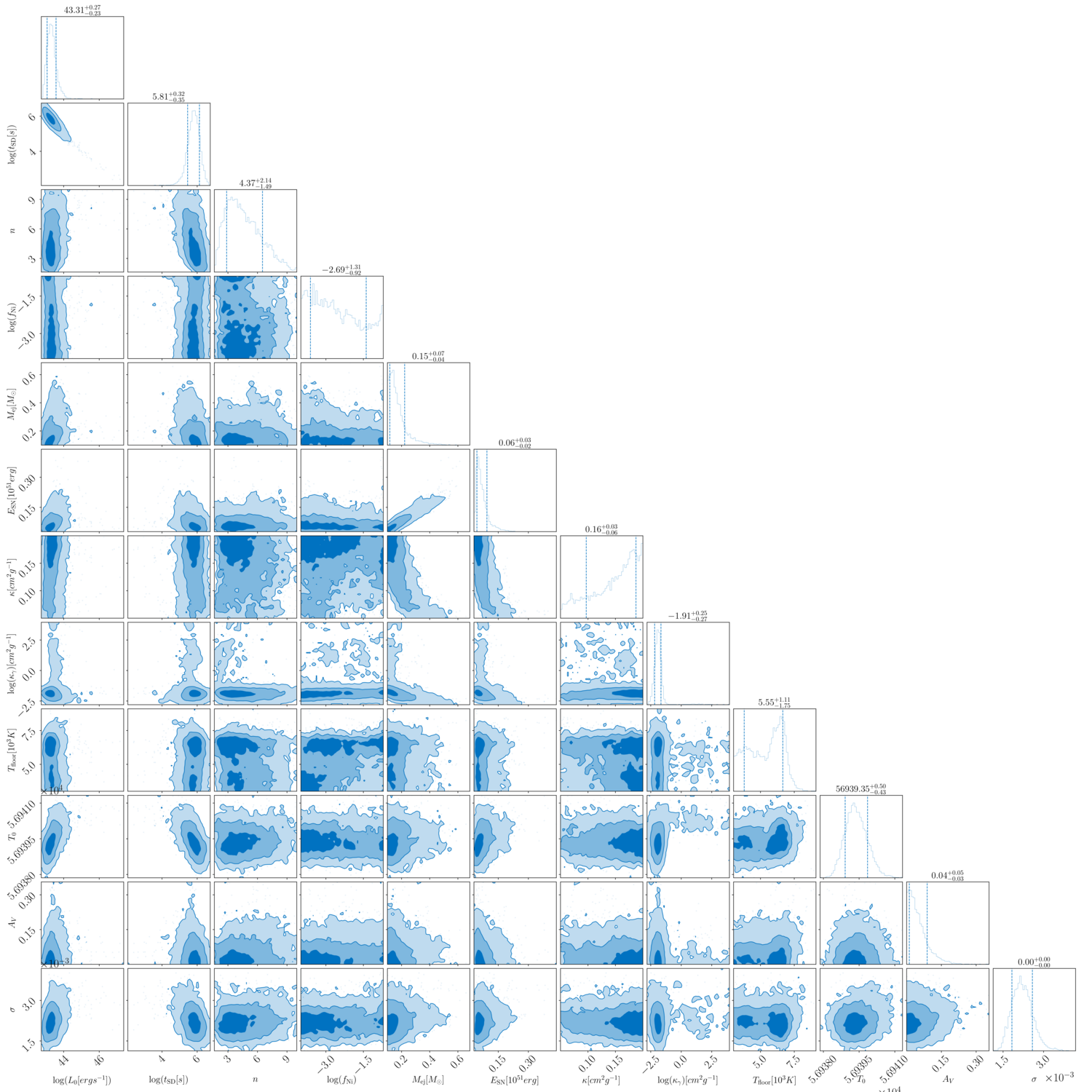}
\caption{Corner plot for iPTF14gqr, a USSN.}%
\label{fig:14gqr_corner}
\end{figure*}

Here we show the full posteriors for the simulated (Figure \ref{fig:simulated_corner}) and real supernovae examined as case studies, SN 2015bn (Figure \ref{fig:15bn_corner}), SN 2007ru (Figure \ref{fig:07ru_corner}), ZTF20acigmel (Figure \ref{fig:camel_corner}), and iPTF14gqr (Figure \ref{fig:14gqr_corner}).

%%%%%%%%%%%%%%%%%%%%%%%%%%%%%%%%%%%%%%%%%%%%%%%%%%

% Don't change these lines
\bsp	% typesetting comment
\label{lastpage}
\end{document}